\documentclass{aa}  

\usepackage{graphicx}
\usepackage{txfonts}
\usepackage[colorlinks]{hyperref}

\def\aap{A\&A}
\def\apj{ApJ}
\def\apjs{ApJS}

\def \hi {\ion{H}{i}}
\def\h2{H$_2$}

\def\kms{km\,s$^{-1}$}

\def\deg{\hbox{$^\circ$}}
\def\arcmin{\hbox{$^\prime$}}

\def\fdg{\hbox{$.\!\!^\circ$}}
\def\farcm{\hbox{$.\mkern-4mu^\prime$}}

\defcitealias{Kalberla2021}{Paper~I}
\defcitealias{Kalberla2023}{Paper~II}

\begin{document}

\title{Caustics and velocity caustics in the diffuse interstellar
  medium at high Galactic latitudes}


\subtitle{}

   \author{P.\ M.\ W.\ Kalberla }

\institute{Argelander-Institut f\"ur Astronomie, University of Bonn,
           Auf dem H\"ugel 71, 53121 Bonn, Germany \\
           \email{pkalberla@astro.uni-bonn.de}
 }

   \authorrunning{P.\,M.\,W. Kalberla } 

   \titlerunning{Caustics in the diffuse ISM }

   \date{Received 25 September 2023 / Accepted 8 December 2023}

  \abstract 
{The far-infrared (FIR) distribution at high Galactic latitudes,
  observed with {\it Planck}, is filamentary with coherent structures in
  polarization. These structures are also closely related to
  \hi\ filaments with coherent velocity structures. There is a
  long-standing debate about the physical nature of these
  structures. They are considered either as velocity caustics,
  fluctuations engraved by the turbulent velocity field  or as cold
  three-dimensional density structures in the interstellar medium (ISM).
}
{We discuss different approaches to data analysis and interpretation in order to
  work out the differences.  }
{We considered mathematical preliminaries for the derivation of caustics
  that characterize filamentary structures in the ISM.  Using the
  Hessian operator, we traced individual FIR filamentary structures in
  \hi\ from channel maps as observed and alternatively from data that
  are provided by the velocity decomposition algorithm (VDA).  VDA is
  claimed to separate velocity caustics from density effects.  }
{Based on the strict mathematical definition, the so-called velocity
  caustics are not actually caustics. These VDA data products may contain caustics
  in the same way as the original \hi\ observations.  Caustics derived
  by a Hessian analysis of both databases are nearly identical with a
  correlation coefficient of 98\%. However, the VDA algorithm leads  to a
  30\% increase in  the alignment uncertainties when fitting
  FIR/\hi\ orientation angles. Thus, the VDA velocity crowding concept
  fails to explain the alignment of FIR/\hi\ filaments at $|b| > 20
  \degr$.  We used \hi\ absorption data to constrain the physical nature
  of FIR/\hi\ filaments and determine spin temperatures and volume
  densities of FIR/\hi\ filaments. \hi\ filaments exist as cold neutral medium
(CNM)
  structures; outside the filaments no CNM absorption is detectable.  }
{ The CNM in the diffuse ISM is exclusively located in filaments with FIR
  counterparts. These filaments at high Galactic latitudes exist as cold
  density structures; velocity crowding effects are negligible.    }


  \maketitle
%

\section{Introduction}
\label{Intro}

Take a teacup. At the bottom of this cup you can   observe a
caustic, the image of the Sun focused typically into two bright curves
with a central peak. Caustics can also be generated by refraction, like
a rainbow or the wiggling patterns at the bottom of a swimming pool. 
  Simple structures of this kind can be explained by ray
  tracing.\footnote[1]{The first known drawing of these structures is by Leonardo
    da Vinci, available as Codex Arundel, folio 86,
    \url{https://www.bl.uk/catalogues/illuminatedmanuscripts/record.asp?MSID=6454}}
  Gravitational lensing and microlensing events are examples of more
  complex cases. The shapes of these caustics depend on the source
  distribution and the intervening matter
\citep[e.g.,][]{Blandford1992}. Caustics are also found in topological
  structures of the cosmic web, such as voids, peaks, walls, and
  filaments \citep[e.g.,][]{Bond1996}. A general formalism for caustics
  that also applies   to such complex structures has been given recently by
  \citet{Feldbrugge2018}.  
  
For the interpretation of caustics it is in general necessary to know
some details of the  physics of the imaging system as well as its
geometry. This implies that some initial model assumptions are
needed. Consent about model conditions is not always granted; different
groups may advocate controversial models, and we are then faced with a
diversity of possible interpretations. In this paper we discuss such a
case, the interpretations of observed caustics in the diffuse
interstellar medium. Prominent filamentary structures, observable with
{\it Plank} at 857 GHz or 353 GHz 
\citep{Planck2016a,Planck2016b,Planck2016c},  are found to be well
aligned with the ambient magnetic field in the diffuse interstellar
medium (ISM).  These  small-scale structures caused by emission from dust
in the far-infrared (FIR) are well aligned with small-scale
\hi\ structures and the discussion is about the nature of the observed
intensity enhancements in \hi\ channel maps.

One the one hand, these filamentary structures have been interpreted as caustics
originating from real physical structures, coherent \hi\ fibers with
local density enhancements in position-velocity space. The ISM is
understood as a multiphase medium with a diffuse warm neutral medium
(WNM) in pressure equilibrium with the embedded   cold neutral medium
(CNM) \citep[e.g.,][]{Wolfire2003}.  This CNM, observed with a
characteristic narrow velocity distribution, is filamentary and
associated with cold dust, stretched out along the magnetic field lines
(e.g., \citealt{Clark2014}, \citealt{Kalberla2016},
  \citealt{Clark2019}, or \citealt{Peek2019}). 

Counterarguments are based on \citet{Lazarian2000}. The observed filaments
 accordingly indicate velocity caustics caused by turbulent velocity
fluctuations. The observed intensity enhancements in channel maps with
small velocity widths originate in this approach from nonlinear density
fields that are transformed by coherent turbulent velocities. The
emission is generated in different volume elements, separated along the
line of sight. Intensity enhancements are thus generated by velocity
crowding. \hi\ intensity enhancement may therefore not be interpreted as
real coherent density structures, but reflect the properties of random
turbulent fields. Such structures are referred to as velocity caustics

In this paper we   discuss the two opposing positions. We focus
on phenomena that lead to observable caustics in the diffuse
interstellar medium. The scientific community needs to 
distinguish between the two different approaches that are described in almost
identical terms, caustics and velocity caustics. In the first case we
have generic caustics ($\kappa \alpha \upsilon \sigma \tau \iota \kappa
\acute{o} \zeta$), as introduced in the first paragraph. These caustics
are deterministic. A particular source distribution generates a
characteristic image, and a trace-back is possible.  In the second case of velocity caustics, intensity structures originate from
turbulent velocity fluctuations.  The statistical nature of this process
does not allow conclusions about distinct source properties like
temperature and volume density of the \hi.

This paper is structured as follows. We first consider in Sect. \ref{Caustic} the morphology of generic
caustics and the mathematical approaches that define such caustics.  Next we
discuss in Sect. \ref{VCA} velocity caustics in context with the
velocity channel analysis (VCA), proposed by \citet{Lazarian2000} and
the derived velocity decomposition algorithm (VDA) by
\citet{Yuen2021}. In Sect. \ref{VDA_caustics} we calculate the proposed
VDA velocity and density contributions, and use the Hessian operator to
determine VDA eigenvectors and orientation angles that are then compared
with previous results from \citet[][hereafter Paper I]{Kalberla2021}.  The
numerical results are discussed in Sect. \ref{Evaluation}; \hi\ power
spectra are discussed  in Sect. \ref{VDA_power}. The physical conditions in caustics
are considered in Sect. \ref{physics}. We demonstrate that
FIR/\hi\ filaments are unambiguously related to the CNM and demonstrate
the differences between velocity slicing and velocity tracing.  After a
summary in Sect. \ref{Summary}, we conclude that FIR/\hi\ filaments can
be traced as small-scale structures in position, associated with coherent
structures in velocity with narrow linewidths. FIR/\hi\ caustics are
density structures in the position-velocity space.

\section{Caustics: A morphological approach }
\label{Caustic}

Caustics are considered by mathematicians as part of the theory of local
singularities of smooth morphisms; physicists are more interested in the
stability of natural systems. In this context we start with a brief
description of the mathematical background and methods that are useful
for the determination and analysis of caustics. 

\subsection{Caustics as Morse complexes }
\label{Catastrophe_theory}

Caustics are defined by their particular topology, and it is challenging
to use a morphology-based data analysis to derive physical properties of
observed intensity structures. The mathematical framework for a
treatment of caustics is based on investigations by \citet{Thom1975} and
\citet{Arnold1985}.  Caustics are usually considered in the more general
framework of catastrophe theory.\footnote[2]{R. Thom received the
  Fields Medal in 1958  for his contributions to the topology of differentiable
  manifolds, and was later the founder of catastrophe theory.} A recent
rigorous treatment of the mathematical background can be found in
\citet{Castrigiano2004}.  Caustics in smooth differentiable maps are
related to critical points, positions where the derivative of the
investigated distribution function is zero. In 2D maps such positions
identify minima, saddle points, and maxima. These are basins, passes, and
peaks if we take mountain shapes as examples. These formations are
usually structurally stable, and this is the property of generic critical
points, also called Morse critical points, that can locally be
approximated by quadratic functions.  These points need to be
 nondegenerate, the determinant of the Hessian (the product of the
eigenvalues) has to be nonzero.  Thus, for our 2D application,
structurally stable critical points are characterized locally by $x^2 +
y^2$ (local minima), $x^2 - y^2$ or $y^2 - x^2$ (saddles), and $-x^2
-y^2$ (local maxima).

Caustics in more than two dimensions are far more complex. Structure
analyses of the cosmic web need to take these general cases into account
and have recently been discussed by \citet{Sousbie2011a},
\citet{Sousbie2011b}, \citet{Feldbrugge2018}, and
\citet{Feldbrugge2019}.\footnote[3]{A comprehensive introduction to
  catastrophe theory can be found at
  \url{https://jfeldbrugge.github.io/Catastrophe-Theory/}} Important
for our 2D application is that generic caustics on the plane can only
exist along   $A_3$ caustics, as classified by Arnold,  or cusps, which is the notation used by Thom (see Sect. 21.3, corollary 3 in
\citealt{Arnold1985}).  In general, the $A_3$ caustics characterize 
  structures that are commonly labeled as filaments or fibers.

Filaments in the ISM have been described by \citet{Schisano2014} as
elongated regions with a   higher brightness contrast with
respect to its surrounding. It is important that visual perceptions of
this kind can be (and need to be) unambiguously verified as 
caustics.  Such regions represent what is considered in the sense of a
general morphological approach as descending manifolds, passes including
connected peaks. Structures of this kind, $A_3$ caustics, can rigorously
be extracted by using a Hessian-based filament identification scheme
with negative eigenvalues. Alternatively to the Hessian matrix,
different methods can be applied. The rolling Hough transform
\citep[RHT,][]{Clark2014} and FILFINDER \citep{Koch2015} have been shown
by \citet{Soler2020} to lead to equivalent results.

Formally, the set of all descending manifolds is usually
referred to as the Morse complex \citep[][definitions 2.4 and
  2.5]{Sousbie2011a}. Filamentary structures can be determined as
surfaces of manifolds associated with critical points of   order one
(saddles) where the Hessian matrix has everywhere exactly one negative
eigenvalue (described also as Morse index one). A peak is of    order
two.\footnote[4]{FIR filaments with a threshold $\lambda_{-} <
  -1.5\ \mathrm{K deg}^{-2}$, discussed in \citetalias{Kalberla2021},
  have at high Galactic latitudes for 3\% of all positions (6\% all sky,
  respectively) a Morse index of two, and for 26\% (39\% all sky) an index
  of one. } Peaks are isolated and are always connected to
passes. Considering these peaks as parts of filaments (otherwise peaks
would disrupt filaments) leads to the constraint that for filaments the
Morse index must be at least one.

As discussed in Sect. \ref{VDA_caustics}, a number of publications since
the year 2000 characterize filamentary structures in the ISM as velocity
caustics in a rather unsharp and ambiguous way. This is in conflict with
the original definition of caustics by \citet{Thom1975} and
\citet{Arnold1985}, and for the sake of clarity  in
the following, in most cases we use the term filaments to  describe $A_3$
  caustics that are characterized by descending manifolds with Morse
  indices of one or two. Velocity caustics, caused by velocity crowding,
  have a filamentary appearance, but may not be considered to originate
  from homogeneous fibers, theads, or filaments.

\subsection{The Hessian operator as applied to FIR and \hi\ data}
\label{Hessian}

Morse complexes can best be identified with the Hessian matrix $H$, which
is based on partial derivatives of the intensity distribution.  This
  method was used for the first time by \citet{Polychroni2013} for the
  identification of filamentary structures in Orion
  A. \citet{Planck2016a} used the Hessian matrix to study the relative
  orientation between the magnetic field and structures traced by
  interstellar dust. A recent application of this method to filamentary
  structures in the Galactic plane is presented in  \citet{Soler2022}. 

In \citetalias{Kalberla2021} and \citet[][hereafter Paper
  II]{Kalberla2023} high Galactic latitude structures in FIR and
\hi\ data were analyzed. Here we briefly summarize this approach.
The Hessian matrix is defined as
\begin{equation}
     \label{eq:hessI} 
        H(x,y)\, \equiv \, \left ( \begin{array}{cc} H_{xx} & H_{xy }\\
            H_{yx} & H_{yy} \end{array} \right ).
\end{equation}  
Here {\it x} and {\it y} refer to the true angles in longitude and
latitude. The second-order partial derivatives are $H_{xx}=\partial^2 I
/ \partial x^2$, $H_{xy}=\partial^2 I / \partial x \partial y$,
$H_{yx}=\partial^2 I / \partial y \partial x$, and $H_{yy}=\partial^2 I
/ \partial y^2$.

The eigenvalues of H, 
\begin{equation}
\label{eq:lambda}
\lambda_{\pm}=\frac{(H_{xx}+H_{yy}) \pm \sqrt{(H_{xx}-H_{yy})^2+4H_{xy}H_{yx}}}{2},
\end{equation}
describe the local curvature of the data (\hi\ and FIR as described in
\citetalias{Kalberla2023}). Eigenvalues $\lambda_- < 0~
\mathrm{K/deg}^{-2}$ with the associated eigenvectors are in direction
of least curvature and indicate filamentary structures, passes, or
ridges. We use negative thresholds for FIR and \hi\ to identify only 
significant descending manifolds, hence caustics. The local orientation of filamentary structures relative to
the Galactic plane (defined by the eigenvectors) is given by the angle
\begin{equation}\label{eq:theta}
\theta =
\frac{1}{2}\arctan\left[\frac{H_{xy}+H_{yx}}{H_{xx}-H_{yy}}\right],
\end{equation}
in analogy to the relation
\begin{equation}\label{eq:theta2}
\theta_S =\frac{1}{2}\arctan \frac{U}{Q},
\end{equation}
from polarimetric observations that provide the Stokes parameters $U$
and $Q$.

The Hessian analysis, including the determination of local orientation
angles along the filaments, was applied in \citetalias{Kalberla2021} and
\citetalias{Kalberla2023} to {\it Planck} FIR data at 857 GHz. Also considered were
HI4PI \hi\ observations \citep{HI4PI2016}, combining data from the
Galactic All Sky Survey (GASS; \citealt{Kalberla2015}), measured with the
Parkes radio telescope and the Effelsberg-Bonn \hi\ Survey (EBHIS;
\citealt{Winkel2016}) with data from the 100 m telescope.

The Hessian operator was applied to {\it Planck} 857 GHz FIR data with a
threshold $\lambda_{-} < -1.5\ \mathrm{K deg}^{-2}$ to ensure the
significance of the eigenvalues. This procedure defines an
ensemble of isolated FIR filaments. In the next step this analysis was
repeated for all \hi\ channels, this time with  an eigenvalue 
threshold $\lambda_{-} < -50\ \mathrm{K deg}^{-2}$.  These limits
  take different sensitivities and intensity scaling in FIR and
  \hi\ into account; we note that the applied thresholds are related to
  observed intensities due to the constant multiple rule for
  derivatives.  Each of the FIR manifolds is found to be associated in
\hi\ with a distribution of filaments in velocity space (represented by
channel maps) that approximately fit  local shapes of the FIR
structures. Orientation angles $\theta$ were used to resolve the
velocity ambiguities by searching for a velocity-space alignment of
\hi\ structures and orientation angles with the FIR emission.

Each individual FIR position within each of the descending manifolds was
compared with each channel of the derived \hi\ manifolds with the aim of
finding the velocity with the best match for the local orientation angle
$\theta$. Best fit results, on average accurate to $\sigma_{\theta} =
4\fdg1 $, were found for narrow velocity channels with the original
instrumental resolution at high Galactic latitudes, see 
\citetalias{Kalberla2021}.  These morphologically derived velocities
define the velocity field of the FIR filaments. Radial velocities along
individual filaments were found to be well defined and coherent with an
average random scatter of $\Delta v_{\mathrm{LSR}} = 5.24 $ \kms\ along
the filaments that was interpreted in \citetalias{Kalberla2023} as
internal turbulent motions of the filaments.

In summary, the Hessian operator was used in \citetalias{Kalberla2021}
to identify Morse complexes, filaments with common properties
for FIR, and \hi. Along these filaments FIR and \hi\ share common
orientation angles at well-defined coherent \hi\ velocities.  Known FIR
distances allow  volume densities to be determined from observed column
densities. The Morse complexes can straightforwardly be interpreted as
cold density filaments (see \citetalias{Kalberla2023}). Minkowski
functionals were found useful to derive aspect ratios and filamentarity
of the FIR/\hi\ filaments.

\section{Probing the velocity channel analysis}
\label{VCA}

Velocity channel analysis   \citep[VCA,][]{Lazarian2000} deals with
emission spectra in individual velocity slices (channel maps) and
  derives dependences on the statistics of turbulent velocity
and density fields by varying the slice thickness. Here we explain
briefly the VCA preliminaries.

\subsection{Velocity channel analysis and velocity caustics}
\label{Vel_caustic}

\hi\ and molecular line observations are usually organized in
position-position-velocity (PPV) data cubes. Channel maps at constant
velocities measure line intensities that depend on the physical
conditions of the gas at a particular observed radial velocity. The derived
parameters are affected by turbulent velocity fluctuations along the
line of sight. Additional fluctuations in density are expected. The
problem of disentangling velocity and density fluctuations was first
addressed in \citet{Lazarian2000}, and subsequently  by
\citet{Lazarian2004,Lazarian2006,Lazarian2008}. These authors represent
the PPV correlation function as a sum of two uncorrelated terms that
depend for each velocity channel either on fluctuations of density or
velocity. Accordingly, narrow velocity slices are supposed to be
dominated by velocity perturbations and  broad slices in the limit are dominated by column
density effects.  VCA aims in this way to disentangle velocity and density
statistics. The claim after these investigations is that structures
observed in narrow velocity intervals are caused predominantly by
velocity caustics.

Generic caustics are associated with singularities of gradient maps. The
most convenient way to search for caustics in 2D is the use of a
Hessian-based filament identification scheme.  Velocity caustics,
generated by velocity crowding, were claimed to have been found in
observations and magnetohydrodynamics (MHD) simulations by a number of authors
(e.g., \citet{Esquivel2003}, \citet{Chepurnov2009}, \citet{Padoan2009},
\citet{Lazarian2018}, and \citet{Ho2023}. However, none of these publications
provides  a rigorous identification of structures as generic
caustics or parts of Morse complexes in the sense of
Sect. \ref{Catastrophe_theory} or as described by \citet{Thom1975} and
\citet{Arnold1985}.  In many cases numerical simulations ignore the
  thermal broadening of the lines that tends to wash out velocity
  fluctuations \citep[see][Sect. 3.1]{Clark2019}. Structures that are visible in
  simulated PPV maps   in these cases are not representative for observed
  PPV maps. 

\subsection{Definition of the velocity decomposition algorithm }
\label{VDA}

Following the  theory by \citet{Lazarian2000}, \citet{Yuen2021}
developed a method for decomposing an observed PPV data cube into two
separate cubes that are supposed to contain independently the turbulent
velocity and density information. The decomposition is based on the
postulate that in case of MHD turbulence the density and the velocity
fluctuations are statistically uncorrelated. Using VDA notations this is
formulated as $p = p_v + p_d$ with $\langle p_v p_d \rangle = 0 $, for
an ensemble average indicated by $\langle ... \rangle$. The observed
channel map\footnote[5]{Observed data $p$ are brightness temperatures
  $T_{\mathrm{B}}$ as weighted averages over the instrumental velocity
  window $\delta v$. For consistency with \citet{Yuen2021}, we use $p$ in
  place of $T_{\mathrm{B}}$.} $p$ is decomposed in its velocity
contribution $p_v$ and density part $p_d$ according to
\begin{equation}
\begin{aligned}
  p_v &= p - \left( \langle pI\rangle-\langle p\rangle\langle I \rangle\right)\frac{I-\langle I\rangle}{\sigma_I^2},\\
  p_d &= p-p_v\\
  &=\left( \langle pI\rangle-\langle p\rangle\langle I \rangle\right)\frac{I-\langle I\rangle}{\sigma_I^2},
\end{aligned}
\label{eq:pvd}
\end{equation}
with $I=\int p(v) dv$ the total \hi\ intensity (or column density)
along the line of sight and $\sigma_I^2 = \langle (I - \langle I
\rangle)^2 \rangle$. The necessary condition is that the velocity width
$\Delta v$ of the PPV channel map is small in comparison to the
effective velocity width of the observed \hi\ gas.  To relate this
condition to the analysis in \citetalias{Kalberla2021} we recall that
  $\Delta v = 1$ \kms\ was used there. Typical cold \hi\ structures have a
velocity width of 3 \kms, and thus   \hi\ is well resolved. 
VDA velocity fluctuations are noticeable only in narrow velocity
intervals, but vanish if velocity slices get thicker; $p_v = 0 $ for
$\Delta v\rightarrow \infty$. Per the definition, VDA density fluctuations
scale as $p_d\propto I$, in particular for low sonic Mach numbers $M_s
\ll 1$ \citep[][Sect. 3]{Yuen2021}.

It is important to realize that Eq. (\ref{eq:pvd}) represents a
decomposition of velocity and density fluctuations (positive or negative
deviations from the mean),  while the observed \hi\ channel maps $p = p_v
+ p_d $ are density-weighted emission profiles. The differences between the two
approaches are explained in Fig. 2 of \citet{Yuen2021}.  VDA densities
$p_d $, although  they are denoted as densities, are not to be
confused with volume densities. This can be seen from
Eq. (\ref{eq:pvd}), where  $p_v$ and $p_d$ have the same units as $p$ from
observations. The brightness temperatures $T_{\mathrm{B}}$ in K   averages
over the width of the velocity channel in \kms. Volume densities demand
a definition of the volume that is occupied by the \hi\ (or modeling an
object considered as a cloud). This definition is not part of the VCA
or VDA concept. Volume densities in the presence of a dominating velocity
crowding effect are in this context undefined.

According to \citet{Yuen2021}, the term $p_v$ in Eq. (\ref{eq:pvd})
represents the velocity contribution to velocity channels and is
referred to as (or defined as)  the velocity caustics contribution. Although they are called velocity caustics, $p_v$ is not the velocity field; it can
be seen only as a proxy for the part of the data that are (according to
VCA/VDA) affected by the velocity field. These velocity caustics
$p_v$ are not to be confused with caustics as defined by
\citet{Thom1975} or \citet{Arnold1985}. To avoid any conflicts with
caustics that are defined as manifolds in 2D,  for $p_v$ we   use in
the following the notation VDA velocity caustics. The observed data $p$,
as well as the contributions $p_v$ or $p_d$ may contain manifolds that
represent Morse complexes, but each $p$, $p_v$, or $p_d$ velocity slice on
its own represents no more than a smooth differentiable map that needs
to be analyzed for the presence of supposed caustics before further
conclusions can be drawn. Such an analysis is not part of VDA.   VCA
also does not attempt any morphological decomposition. 

\section{Hessian analysis of the VDA databases $p_v$ and $p_d$ }
\label{VDA_caustics}

Following Eq. (\ref{eq:pvd}), we generate $p_v$ and $p_d$ for the
observed $p$ database used in \citetalias{Kalberla2021}. We repeat the
complete Hessian analysis described in Sect. 2 of
\citetalias{Kalberla2021} independently for VDA derived velocity $p_v$
and density $p_d$ structures without modifying any of the program
parameters. In the following we discuss the properties derived from
Hessian eigenvalues and eigenvectors independently.

\begin{figure}[htp] 
   \centering
   \includegraphics[width=9cm]{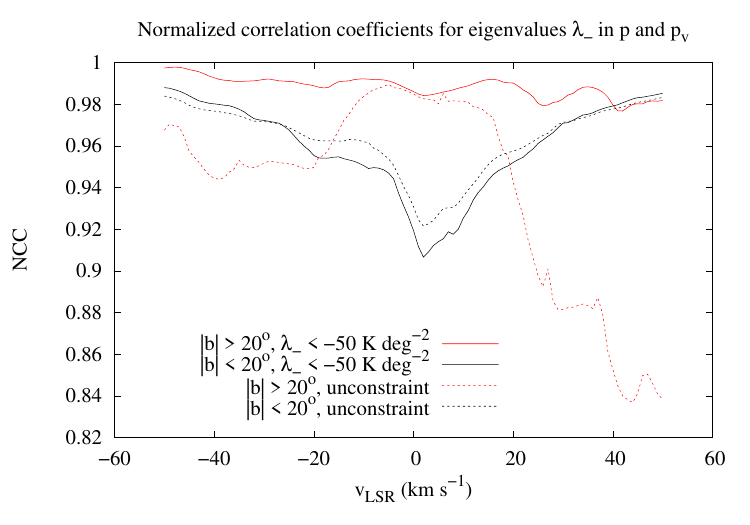}
   \caption{Normalized correlation coefficients for eigenvalues in
     observed intensity $p$ and VDA velocity caustics $p_v$ according to
     Eq. (\ref{eq:pvd}) at high and low Galactic latitudes. Data constrained by $\lambda_{-} < -50\ \mathrm{K
       deg}^{-2} $ (used in \citetalias{Kalberla2021} to compare
     \hi\ with FIR filaments) and unconstrained eigenvalues are distinguished. }
   \label{Fig_NCC}
\end{figure}

\subsection{Correlations of $\lambda_{-}$ eigenvalues in  $p$ and $p_v$ }
\label{VDA_caustics_2}

Inspecting derived eigenvalues for $\lambda_{-}$ in $p$ and $p_v$
visually, we find a close agreement between the two distributions. 
The correlation between the $A$ and $B$ distributions can, according to
\citet{Yuen2021}, be best verified by using the normalized covariance
coefficient
\begin{equation}
  NCC(A,B) = \frac{\langle (A-\langle A\rangle)(B-\langle B\rangle)\rangle}{\sigma_A\sigma_B}
  \label{eq:NCC}
\end{equation}
to characterize correlations between the two 2D maps $A$ and $B$; here we
use the same notations as \citet{Yuen2021}. This measure, also known as
Pearson product-moment correlation coefficient, is scale-invariant and
results in $NCC(A,B)\in[-1,1]$. The case $NCC(A,B) = 0 $ implies that
the two maps are statistically uncorrelated. A perfect correlation
requires that $NCC(A,B) = \pm 1$; the sign reflects the slope of the
linear regression that can be fitted in this case.

The correlation coefficients $NCC(\lambda_{-}(p),\lambda_{-}(p_v))$ for
the eigenvalue distributions at all velocity channels $ -50 <
v_{\mathrm{LSR}} < 50 $ \kms\ are shown in Fig. \ref{Fig_NCC}. The best
correlations with $NCC(\lambda_{-}(p),\lambda_{-}(p_v)) \ga 0.98$ are
found at latitudes $|b| > 20 \degr$ in the case of filamentary structures
constrained in \hi\ by $\lambda_{-} < -50\ \mathrm{K deg}^{-2} $. This is
the same condition as used in \citetalias{Kalberla2021} to ensure that
\hi\ filaments are unaffected by observational
uncertainties. Unconstrained $\lambda_{-}(p)$ and $\lambda_{-}(p_v)$
eigenvalue distributions at high latitudes show only a perfect
correlation for $|v_{\mathrm{LSR}}| \la 8 $ \kms; this is the velocity
range with dominant CNM filaments (see Fig. 11 in
\citetalias{Kalberla2021}).  The $\lambda_{-}(p)$ and $\lambda_{-}(p_v)$
eigenvalue distributions at low Galactic latitudes suffer from confusion
(\citetalias{Kalberla2023}, Sect. 3.1) and the correlation
$NCC(\lambda_{-}(p),\lambda_{-}(p_v))$ degrades significantly,
 whether they are constrained or not.

\begin{figure*}[htb] 
  \centering
  \includegraphics[width=9cm]{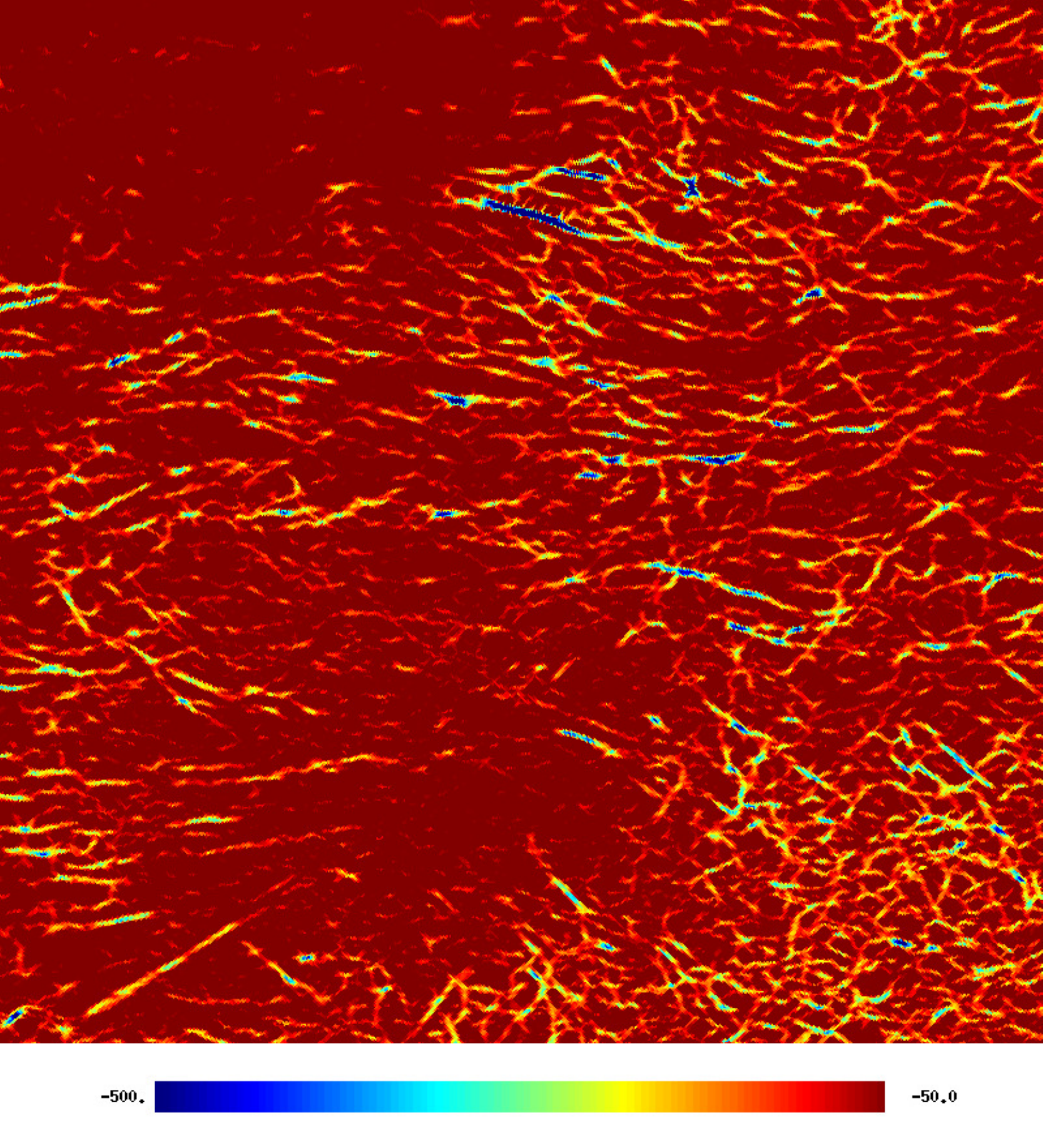}
   \includegraphics[width=9cm]{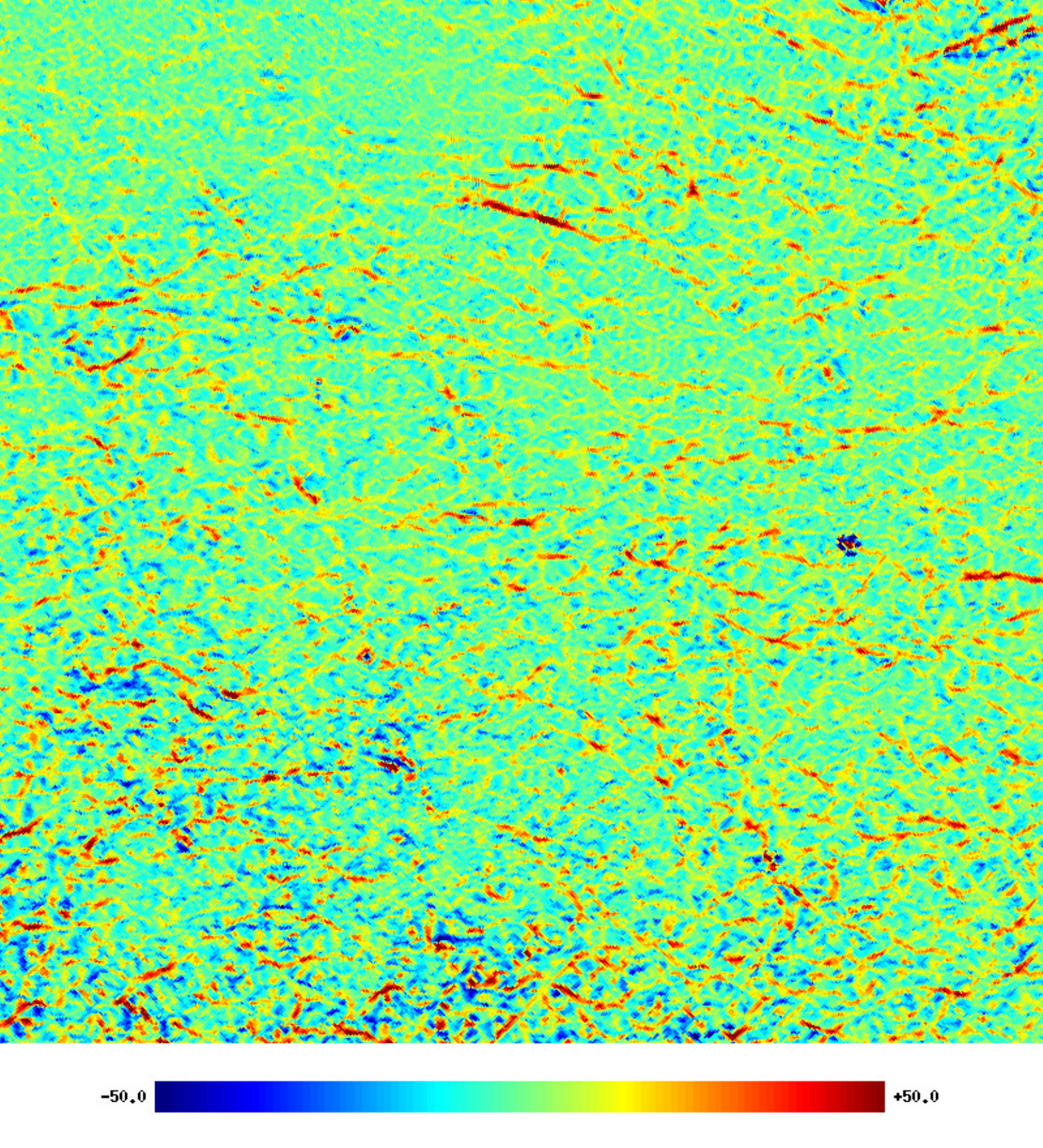}
   \includegraphics[width=9cm]{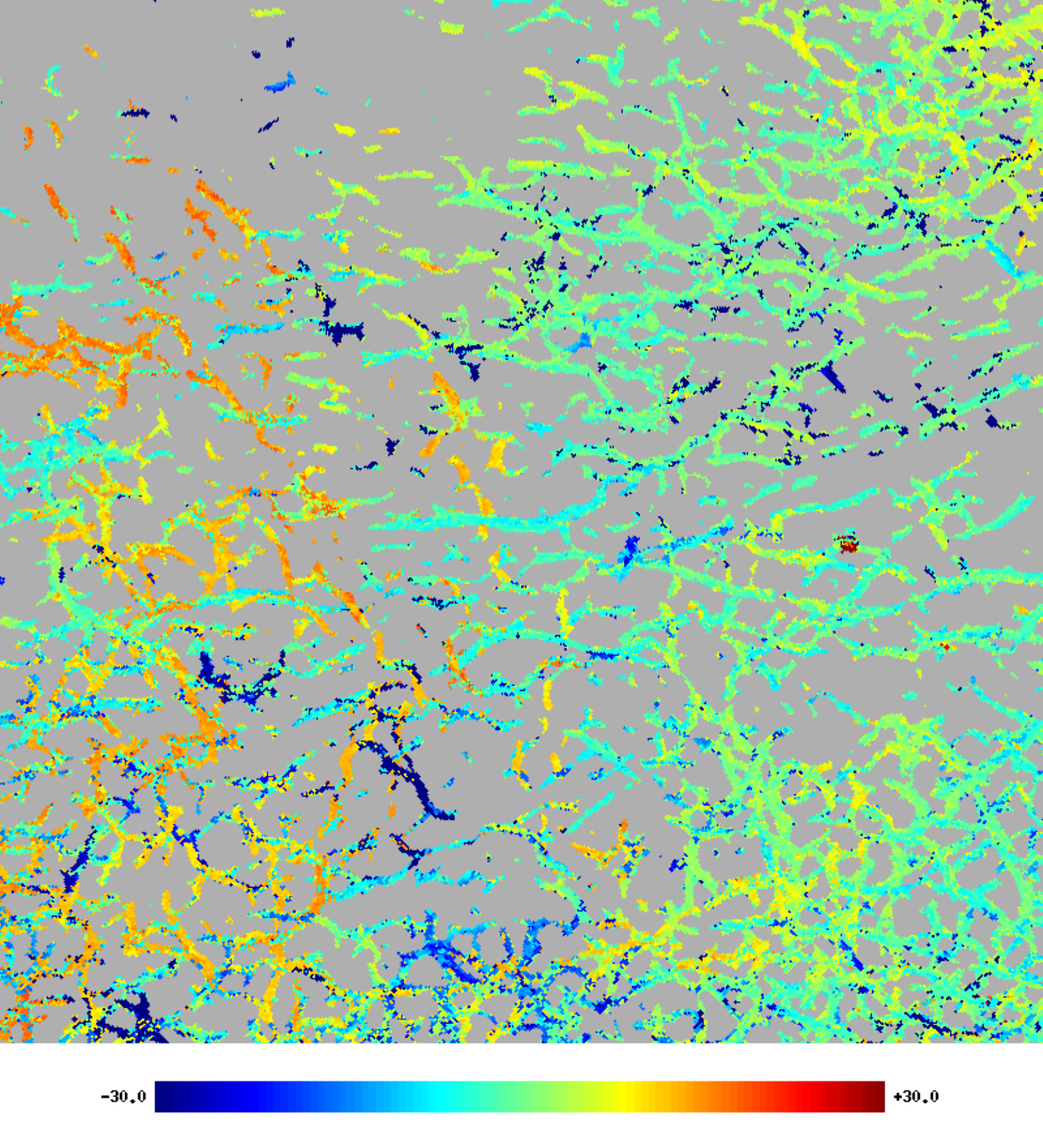}
   \includegraphics[width=9cm]{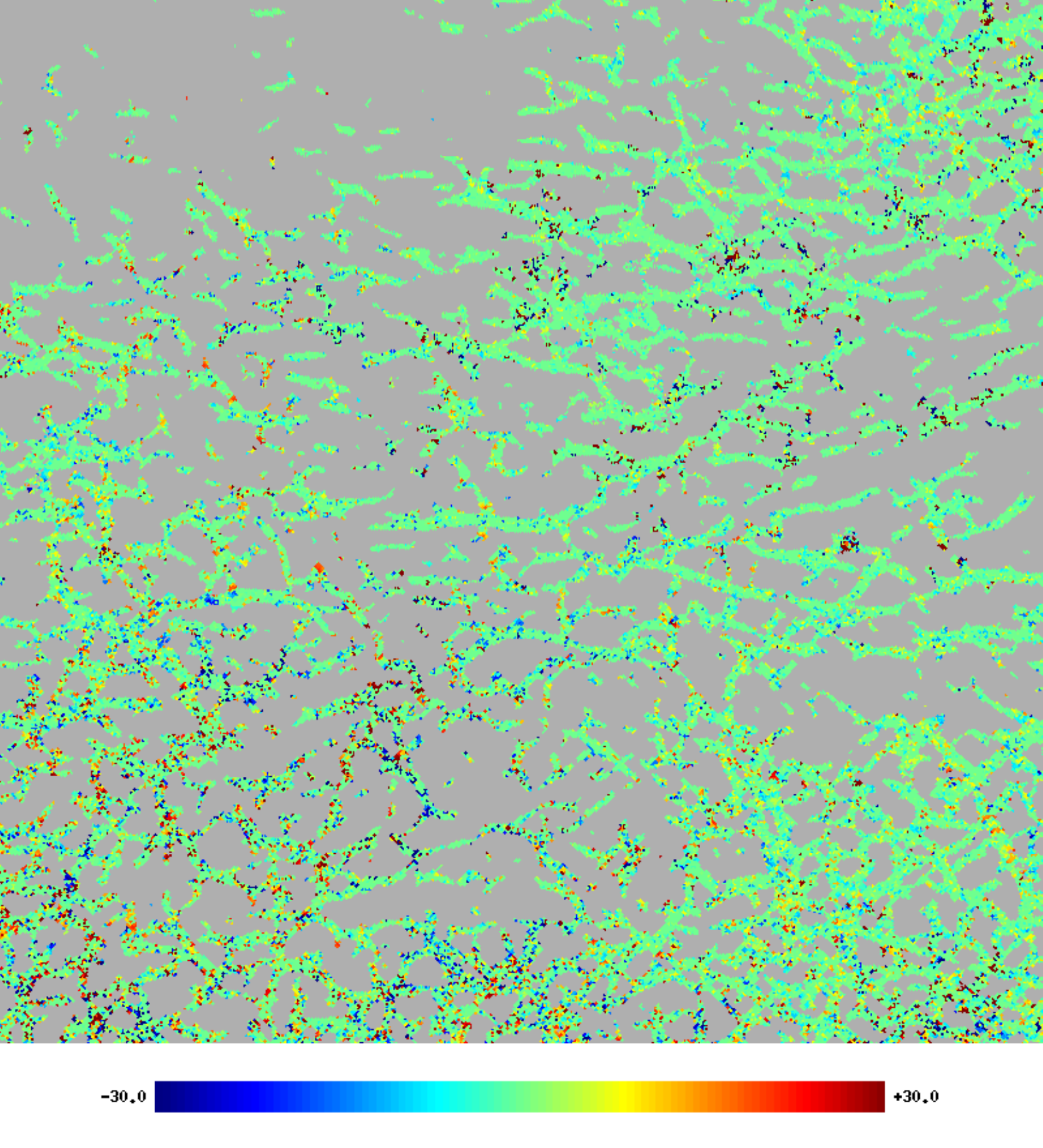}
   \caption{Display of example data in gnomonic projection.
     The field center is at $ l = 160\deg$, $b = 30\deg $, the field
     size 27\fdg7. Top left: eigenvalues $\lambda_{-}(p_v)$ of VDA velocity
     caustics $p_v$ at $v_{\mathrm{LSR}} = 0 $ \kms. Top right:
     deviations $\lambda_{-}(p_v) - \lambda_{-}(p)$ between eigenvalues
     from VDA velocity caustics $p_v$ and observed intensities $p$.
     Bottom left: velocities $v_{\mathrm{LSR}}(p_v)$ of filamentary FIR
     structures in case of \hi\ filaments derived for VDA velocity
     caustics $p_v$. Bottom right: deviations $v_{\mathrm{LSR}}(p_v) -
     v_{\mathrm{LSR}}(p)$ from the velocity field $v_{\mathrm{LSR}}(p)$ shown in Fig. 6 of
     \citetalias{Kalberla2021}. }
   \label{Fig_NoiseMaps_VDA}
\end{figure*}

After discussing the general correlation analysis on Galactic scales we
display on top of Fig. \ref{Fig_NoiseMaps_VDA} an example for the
eigenvalue distribution $\lambda_{-}(p_v)$ together with deviations
$\lambda_{-}(p_v) - \lambda_{-}(p)$. Minor deviations are visible, they
reflect the result that the eigenvalues $\lambda_{-}$ for $p$ and $p_v$
at high Galactic latitudes are highly similar. At the bottom of
Fig. \ref{Fig_NoiseMaps_VDA} we display for comparison the velocity
field of the FIR filaments derived from $p_v$, together with deviations
from the velocity field published in \citetalias{Kalberla2021}.

The Hessian analysis of the VDA velocity term $p_v$, using exactly the
same procedures as in \citetalias{Kalberla2021}, recovers in 97\% of all
case positions in the Morse complexes that are identical with the
positions of the filamentary structures found in
\citetalias{Kalberla2021}. Less than half (43\%) of the previously
derived filament positions have identical VDA filament velocities but 
for 53\% of them the differences are below 1 km/s. Allowing
uncertainties of $\Delta v \la 6 $ \kms\ for the filament velocities
(see Sect. 2.6 of \citetalias{Kalberla2021}), we obtain compatible velocities for 68\% of all filament
  positions. A close
inspection of the lower part of Fig. \ref{Fig_NoiseMaps_VDA} reveals
that positions with significant differences in the derived velocities
are located predominantly on the filament outskirts. In comparison to
the filament centers, these positions have a lower signal-to-noise ratio
(S/N). We also note an increase in the velocity uncertainties toward the
Galactic plane, explainable with increasing confusion due to the
increasing complexity of the \hi\ emission.

\begin{figure*}[htp] 
   \centering
   \includegraphics[width=8.2cm]{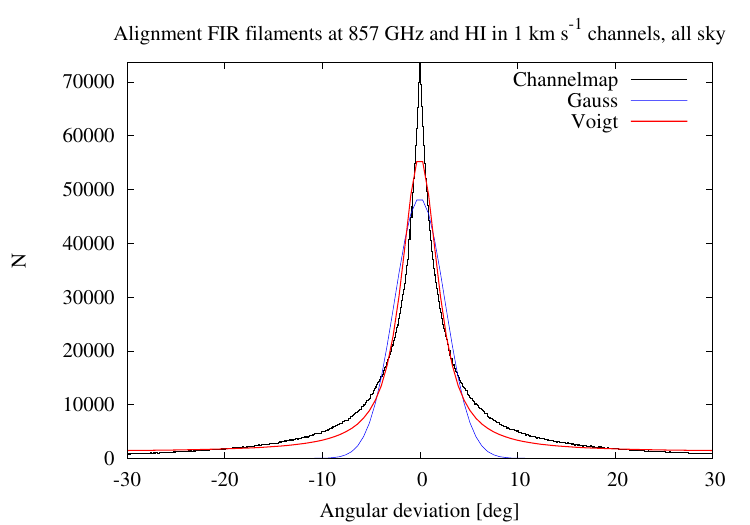}
   \includegraphics[width=8.2cm]{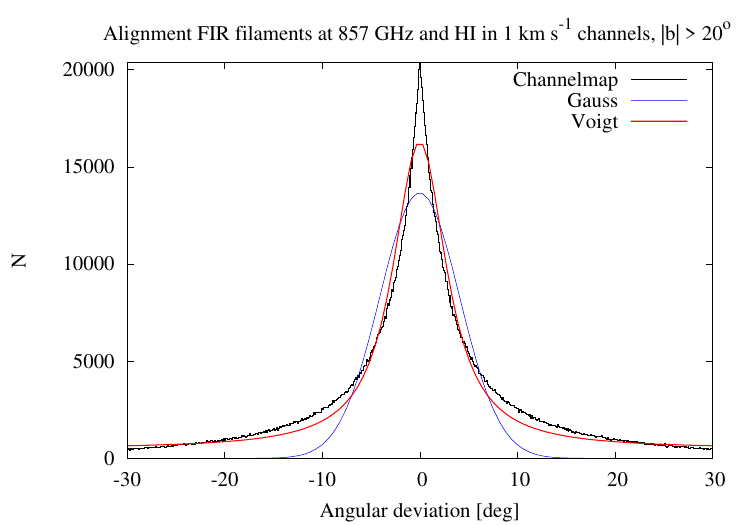}
   \includegraphics[width=8.2cm]{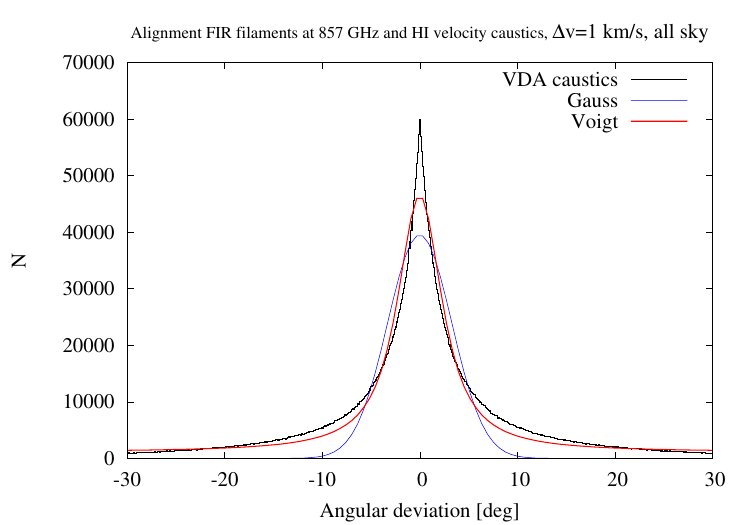}
   \includegraphics[width=8.2cm]{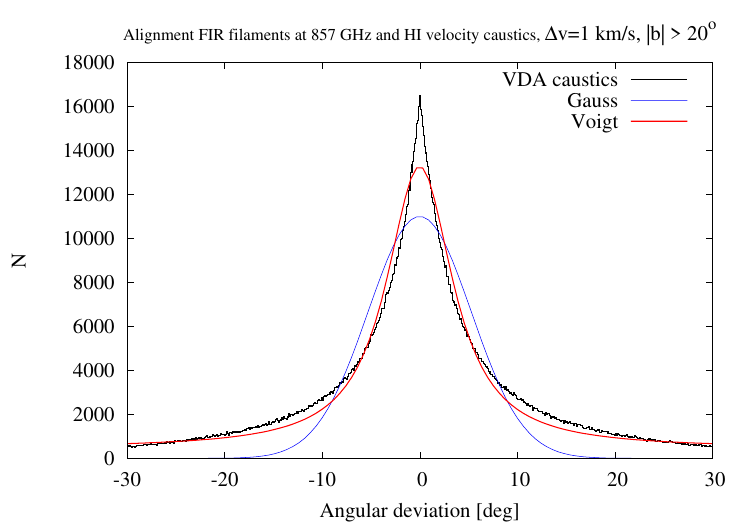}
   \includegraphics[width=8.2cm]{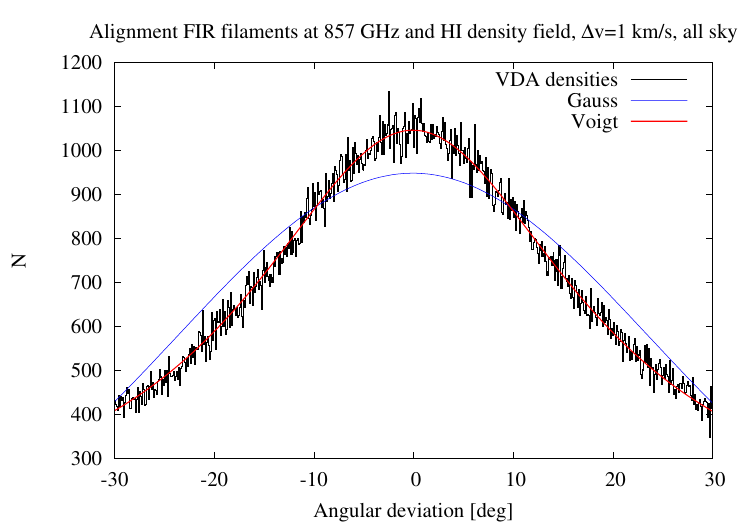}
   \includegraphics[width=8.2cm]{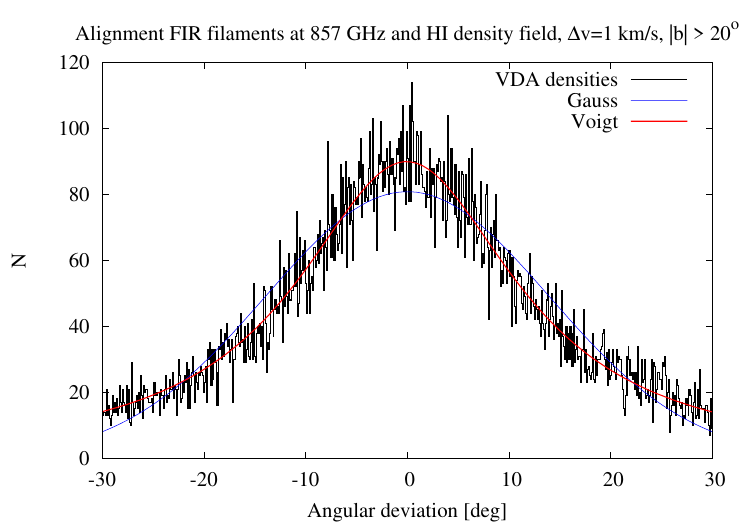}
   \caption{Histograms of angular alignment deviations according to
     Eq. (5) in \citetalias{Kalberla2021} for filamentary
     structures. Top left: {\it Planck} 857 GHz compared with best fit
     single-channel \hi\ filaments, all sky. Top right: {\it Planck} 857
     GHz compared with best fit single-channel \hi\ filaments, $|b| > 20
     \degr$. These two plots were replicated from Fig. 3 in
     \citetalias{Kalberla2021}.  Center left: {\it Planck} 857 GHz
     compared with best fit filaments from the VDA velocity field $p_v$,
     all sky. Center right: {\it Planck} 857 GHz compared with best fit
     filaments from $p_v$ with $|b| > 20 \degr$. Bottom left: {\it Planck}
     857 GHz compared with best fit filaments from the VDA density
     distribution $p_d$, all sky. Bottom right: {\it Planck} 857 GHz
     compared with best fit filaments from the VDA density distribution
     $p_d$, $|b| > 20 \degr$. }
   \label{Fig_Aligne_VDA}
\end{figure*}

\begin{table*}
\caption{Filamentary alignment measures, for comparison with
  Table 1 in \citetalias{Kalberla2021} }             
\label{table:A}      
\centering          
\begin{tabular}{c c c c c c c c c c }     
\hline\hline       
Data 1 & Data 2 & latitude & $f$ & \multicolumn{2}{c}{$\delta \theta$}  & $\xi$ & PRS & $\sigma_{\mathrm{PRS}}$ & Fig. \\ 
{\it Planck} 857 GHz & HI4PI & range & & $\sigma_{\mathrm{Gauss}}$ & $\sigma_{\mathrm{Voigt}}$ & & & \\
\hline                    
18\arcmin\ FWHM & \hi\ in 1 \kms\ channels &  all & .37 & 2\fdg6 & 1\fdg9
& 0.94 & 2873.67 & .17 & \ref{Fig_Aligne_VDA} top left\\  
18\arcmin\ FWHM &  VDA velocity field $p_v$ &  all & .37 & 3\fdg3 &
2\fdg5 & 0.94 & 2856.50 & .18 &  \ref{Fig_Aligne_VDA} middle left \\  
18\arcmin\ FWHM &  VDA density field $p_d$ &  all & .04 & 23\fdg8 & 20\fdg2
& 0.76 & 786.92 & .38 &  \ref{Fig_Aligne_VDA} bottom left \\
\hline
\hline
18\arcmin\ FWHM & \hi\ in 1 \kms\ channels &  $|b| > 20 \degr$ & .24 &
4\fdg1 & 3\fdg1 & 0.92 & 1817.64 & .23 & \ref{Fig_Aligne_VDA} top right \\
18\arcmin\ FWHM & VDA velocity field $p_v$ &  $|b| > 20 \degr$ & .23 &
5\fdg3 & 3\fdg9 & 0.91 & 1789.17 & .24 & \ref{Fig_Aligne_VDA} middle right \\
18\arcmin\ FWHM &  VDA density field $p_d$ &  $|b| > 20 \degr$ & .04 & 14\fdg0 &
13\fdg1 & 0.84 & 207.28 & .31 & \ref{Fig_Aligne_VDA} bottom right\\
\hline   
\end{tabular}
\end{table*}
%

\subsection{Orientation angle alignments between $\theta_{p_v}$,
  $\theta_{p}$, and  $\theta_{\mathrm{FIR}}$  }
\label{VDA_caustics_3}

In the previous subsection we considered agreements in eigenvalues
$\lambda_{-}$. In the following we     take the eigenvectors into
account. These determine the local orientation of filamentary structures
and are conveniently parameterized by the angle $\theta$ according to
Eq. (\ref{eq:theta}). We complete the comparison between filamentary
structures in $p_v$ and $p$ with a summary of orientation angle
alignments in Table \ref{table:A} and Fig. \ref{Fig_Aligne_VDA}. To
allow an easy comparison, we replicate the previously derived best fit
entries from Fig. 3 and Table 1 of \citetalias{Kalberla2021}.

For a detailed definition of the different alignment measures in Table
\ref{table:A} we refer to \citetalias{Kalberla2021}. Here we provide
only a brief description. To measure the angular alignments between two data
sets we calculate first the angular difference between the orientation
angles $\Theta_1$ and $\Theta_2$ at each position:
\begin{equation}
\delta \theta = \frac{1}{2}
\mathrm{arctan}\left[\frac{\mathrm{sin}(2\theta_{1})\mathrm{cos}(2\theta_{2})
    - \mathrm{cos}(2\theta_{1})\mathrm{sin}(2\theta_{2})
  }{\mathrm{cos}(2\theta_{1})\mathrm{cos}(2\theta_{2}) +
    \mathrm{sin}(2\theta_{1})\mathrm{sin}(2\theta_{2}) }\right]
\label{eq:angdif}
.\end{equation} The width of the resulting $\delta \theta$ distribution,
shown in Fig. \ref{Fig_Aligne_VDA}, can be measured by fitting either a
Gaussian or a Voigt function; the Voigt function approximates noisy
data best. Table \ref{table:A} lists the fitted $\delta \theta$
dispersions $\sigma_{\mathrm{Gauss}}$ and $\sigma_{\mathrm{Voigt}}$. The
mean degree of alignment, $\mathrm{\xi} = \left< \mathrm{cos} \phi
\right>$, can be defined for $\phi = 2 \delta \theta$ \citep{Clark2019b}.
A different metric, the projected Rayleigh statistic (PRS), was
calculated according to \citet{Jow2018},
\begin{equation}
\mathrm{PRS} = \sqrt{\frac{2}{N}} \sum_i \mathrm{cos}\, \phi_i,
\label{eq:PRS}
\end{equation}
and the uncertainty of this measure is estimated as 
\begin{equation}
\sigma^2_{\mathrm{PRS}} = \frac{2 \sum_i \mathrm{cos}^2\, \phi_i -
  (PRS)^2 } {N}.
\label{eq:sig_PRS}
\end{equation}
Equation (\ref{eq:PRS}) approximates the more general Eq. (4) of
\citet{Soler2022} in the case of constant statistical weights. This is
justified to a high degree for the all-sky surveys in FIR and
\hi\ that we are using here.

The distributions of angular alignment deviations according to Eq. (5)
of \citetalias{Kalberla2021} between orientation angles of FIR filaments
and \hi\ structures are very similar if we use the VDA $p_v$ in place of
observed \hi\ brightness temperatures $p$. The best fit result
for the velocity width of the slices used, presented in Sect. 2
of \citetalias{Kalberla2021}, remains valid and is obtained when using
\hi\ channel maps with a velocity width of $\Delta v = 1$ \kms.
Filamentary structures in the VDA velocity caustics $p_v$ have a $\sim
30$\% broader $\delta \theta$ distribution compared to the best fit
distribution obtained from observed \hi\ channel maps $p$ (see the
widths of the distributions in Fig. \ref{Fig_Aligne_VDA} and the
dispersions $\sigma_{\mathrm{Gauss}}$ and $\sigma_{\mathrm{Voigt}}$ in
Table \ref{table:A}).  The angular alignment deviations for the VDA
density distribution $p_d$ (Fig. \ref{Fig_Aligne_VDA} bottom) are
unacceptably large, and the filling factors $f = 0.04$ in Table
\ref{table:A} are extremely low. The sky contains a negligible amount of
filaments from the VDA $p_d$ distribution.  The VDA densities $p_d$ are
derived from scaled \hi\ column
densities according to Eq. (\ref{eq:pvd}), and we found already in Sect. 2 of \citetalias{Kalberla2021}
that integrating the \hi\ distribution does not lead to an improvement
in the angular alignment measures.

\section{Evaluation: Caustics in $p_v$ contra $p$  }
\label{Evaluation}

After an introduction to morphological concepts that define caustics
according to \citet{Thom1975} or \citet{Arnold1985}, we discussed
methods to derive caustics. We also introduced VDA, derived the
databases according to the VDA technique and calculated caustics of the
VDA databases. These results were compared with caustics from
\hi\ data in conventional PPV databases. In both cases we correlated FIR
caustics with \hi\ caustics. Based on these preliminaries,  in
the following we discuss the mathematical issues and controversial
interpretations with respect to the physical nature of the
FIR/\hi\ filaments.

\subsection{Velocity   caustics $p_v$ are not caustics}
\label{nocaustics}

The paper by \citet{Yuen2021} specifies in detail how the concept of VCA
velocity caustics can be applied numerically to observations (see
Sect. \ref{VDA}). With with this definition of VDA velocity caustics
$p_v$  for the first time it becomes possible to elaborate the VCA concept
of velocity caustics in detail. Section \ref{VDA_caustics} shows that
VDA velocity caustics $p_v$ according to Eq. (\ref{eq:pvd}), despite
their name, are not caustics as defined by \citet{Thom1975} or
\citet{Arnold1985}. To be clear, structures have been defined as
velocity caustics without any concern about the verification of these
structures as caustics (descending manifolds) in the usual topological
meaning.

In a   way similar to the observed distribution $p$, the VDA descendants 
$p_v$ and $p_d$ may contain morphological structures that can be
identified as caustics. These particular subsets (or manifolds that can
be considered   $A_3$ caustics) need to be determined according to the
restrictions explained in Sect. \ref{Catastrophe_theory}. For this
purpose we applied the Hessian operator, and we   show in
Sect. \ref{VDA_caustics} that $p$ and $p_v$ contain essentially
identical filamentary structures. However, when matching filamentary
\hi\ structures to FIR  caustics, the alignment of VDA $p_v$ structures 
is less accurate by 30\%.

\subsection{Caustics in $p$ and $p_v$ as local diffeomorphisms }
\label{diffeomorphisms}

Caustics in 2D maps are descending manifolds, associated in case of the
observed FIR/\hi\  filamentary structures predominantly with local structures of index
one (Sect. \ref{Catastrophe_theory}). According to the Morse lemma, these
regions can locally be approximated by quadratic functions. These local
coordinate transformations describe local diffeomorphisms
\citep{Castrigiano2004}. Generic critical points are structurally
stable, allowing a smooth transition and approximation by quadratic
functions. Essential for such an approximation in 2D is that the Hessian
matrix has to be nondegenerate. The index of a nondegenerate critical
point fully classifies the critical point up to coordinate
transformations.  The Hessian, Eq. (\ref{eq:hessI}), is determined from
second derivatives and describes the local curvature. Inspecting
Eq. (\ref{eq:pvd}), it is easy to see that $p_v$ and $p_d$ are derived
from $p$ by smooth linear transformations, essentially by subtracting
mean intensities and rescaling.  The terms $ \langle pI\rangle $ and
$\langle p\rangle\langle I \rangle$ in Eq. (\ref{eq:pvd}) are constants,
applied to the scaling factor $(I-\langle I\rangle) / \sigma_I^2$.
Linear transformations as in Eq. (\ref{eq:pvd}) do not affect the Morse
index significantly. Thus, common Morse complexes derived for $p$ and
$p_v$ represent local diffeomorphisms, explaining the high correlation
coefficients $NCC(\lambda_{-}(p),\lambda_{-}(p_v)) \ga 0.98$ (see
Fig. \ref{Fig_NCC}) at high Galactic latitudes.

The concept of local diffeomorphisms simply implies that the transition
from $p$ to $p_v$ is defined by a smooth transformation under the
condition that the two maps remain differentiable. In other words, two
singularities are considered equivalent if and only if there exists a
local transformation that maps them into each other. This applies to
97\% of the caustics in $p$ that replicate as local diffeomorphisms in
$p_v$.

Understanding Morse complexes, derived from $p$ and $p_v$ databases, as
local diffeomorphisms, the assertion that the VDA could derive a
completely new set of unexplored velocity caustics data from every
spectroscopic data set \citep[][Sect. 11, item 9]{Yuen2021} needs to be
questioned. The results from Sect. \ref{VDA_caustics}, summarized in
Table \ref{table:A} and Fig. \ref{Fig_Aligne_VDA}, indicate that the
best alignment between FIR and \hi\ filaments exists for the original
unmodified brightness temperatures $p$. The correlation between
filaments in FIR and VDA velocity caustics $p_v$ is weaker, indicating
that the VDA algorithm does not quite match  the physics of the
filamentary ISM.

\section{VDA power distributions: $P_v$ and $P_d$ contra $P$ }
\label{VDA_power}

\begin{figure}[thp] 
   \centering
   \includegraphics[width=9cm]{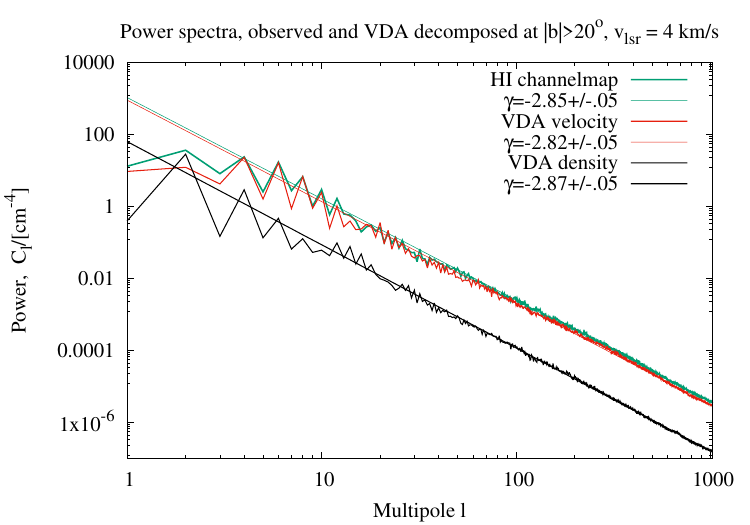}
   \caption{Turbulent power spectra for the observed \hi\ emission $p$
     at high Galactic latitudes at a velocity of $ v_{\mathrm{LSR}} = 4
     $ \kms\ compared with power spectra derived from VDA velocity and
     density contributions $p_v$ and $p_d$.}
   \label{Fig_power}
\end{figure}

A fundamental VCA postulate is that turbulent density and velocity
fields are statistically uncorrelated for the case of MHD turbulence
\citep[][Sect. 6.3.1, Appendix B]{Lazarian2000}. For purely
  isothermal turbulent media global density-velocity correlations are
  unexpected. \citet{Federrath2015}   show  however that for
  non-isothermal multiphase media the Mach number correlates with the
  gas density (their Fig. 7). Understanding the Mach number as a proxy
  for turbulent velocities, one expects, depending on the equation of
  state, that the statistics of velocity and density fluctuations are
  not independent. The VDA algorithm, Eq. (\ref{eq:pvd}), depends by
  construction on the mathematical definition $\langle p_v p_d \rangle =
  0 $ regardless of any actual physics. It is highly misleading to present
  this relation as an observational fact, as is done by \citet{Hu2023} in
  their Appendix A.
  
  Assuming independence of density and velocity, VDA aims to
separate the contributions arising from density and velocity fluctuations
with the corresponding power spectra $ P = P_v + P_d$. For an evaluation
of systematical differences in the power spectra we calculate the
averages of $P_v$ and $P_d$ at high Galactic latitudes. We also relate these
contributions   to the power $P$ from the raw observations.
According to \citet{Lazarian2000} such power spectra should be steep
with power law indices $\gamma < -3$. Power law indices derived from
channel maps are velocity dependent. To match VCA expectations, we  
therefore use the steepest power spectra at a velocity of $ v_{\mathrm{LSR}}
= 4 $ \kms\ derived from HI4PI data (see \citealt{Kalberla2019},
Fig. 9). For the VDA velocity power spectrum $P_v$ we obtain a power law
index $\gamma_{\mathrm{v}} = -2.82 \pm 0.05 $, almost identical with
$\gamma_{\mathrm{obs}} = -2.85 \pm 0.05 $ from the observed \hi\ channel
map at the same velocity (see Fig. \ref{Fig_power}). The spectral index
$\gamma_{\mathrm{d}} = -2.87 \pm 0.05 $ for the VDA density is by
definition velocity independent and identical to the power law index of
the total column density distribution. In summary,
Fig. \ref{Fig_power} implies that the spectral indices for $P$, $P_v$,
and $P_d$ are, within the uncertainties, identical. All power spectra are
straight and do not show any bending in contrast to different possible
model contributions in $P_v$ and $P_d$ (see the case studies in Figs. 6
and 7 of \citealt{Lazarian2000}). The predicted systematical spectral index
changes between $P_v$, $P_d$, and $P$ (their Table 1) are not observed.

Velocity decomposition algorithm density fluctuations are marginal and provide only 4\% of the
observed power. This low level is within the uncertainties consistent
with the previous finding (Sect. \ref{VDA_caustics}) that Morse
complexes of $p_v$ recover 97\% of the positions in Morse complexes of
$p$.  Fig. \ref{Fig_power} shows that the power distributions for $p$
and $p_v$ are almost identical.  The derived spectral index $\gamma \sim
-2.85 $ is representative of individual probes of the ISM at high
Galactic latitudes (see Fig. 9 in the review by
\citealt{Naomi2023}). An average index of $\gamma \sim -2.85 \pm 0.02$
  has also been derived by \citet{Mittal2023} after correcting the
  \hi\ distribution for differences in \hi\ path lengths along the line
  of sight (see their Fig. 8). Under the VCA constraint that the
velocity with the steepest power spectrum should be selected, the
decomposition of the observed brightness temperature distribution in VDA
velocity caustics $p_v$ and densities $p_d$ does not alter the observed
spectral index from $P$. This result is consistent with the
investigations in Sect. \ref{VDA_caustics}.  Statistical similarities in
$P$ and $P_v$ reflect general morphological similarities in $p$ and
$p_v$.

\section{The physics of FIR/\hi\ Morse complexes }
\label{physics}

The data that are analyzed in \citetalias{Kalberla2021} suggest
that filamentary FIR and \hi\ structures, observable at high Galactic
latitudes, were shaped by a Galactic small-scale dynamo. This
conclusion is based on the filament curvature distribution along the
filaments (\citetalias{Kalberla2021}, Sect. 4). Furthermore, individual
filament-like structures share common morphological properties.
Filamentary structures are not individuals; they share a network of
filaments. The resulting distribution of aspect ratios ${\cal A}$ and
filamentarities is well defined and continuous without clear upper
limits in ${\cal A}$ \citepalias{Kalberla2023}. The question arises
whether common morphological properties arise from common physical
conditions. The FIR filament emission is broadband, but in \hi\ the
structures are best reproduced in narrow velocity intervals.

\begin{figure}[th] 
   \centering
   \includegraphics[width=9cm]{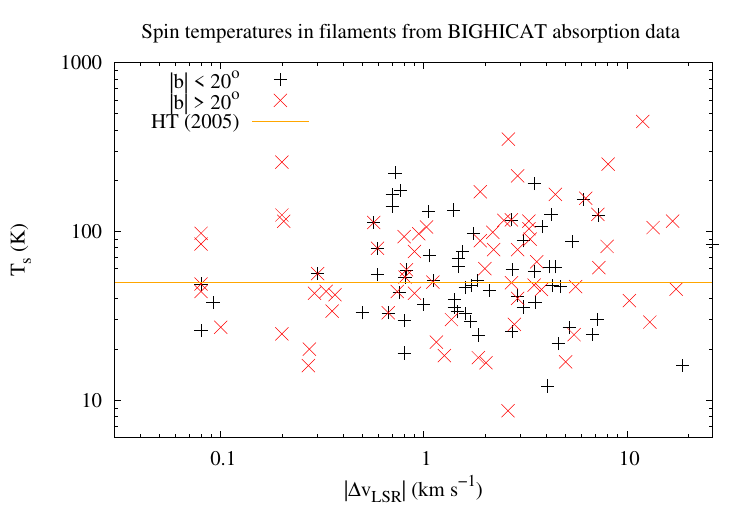}
   \includegraphics[width=9cm]{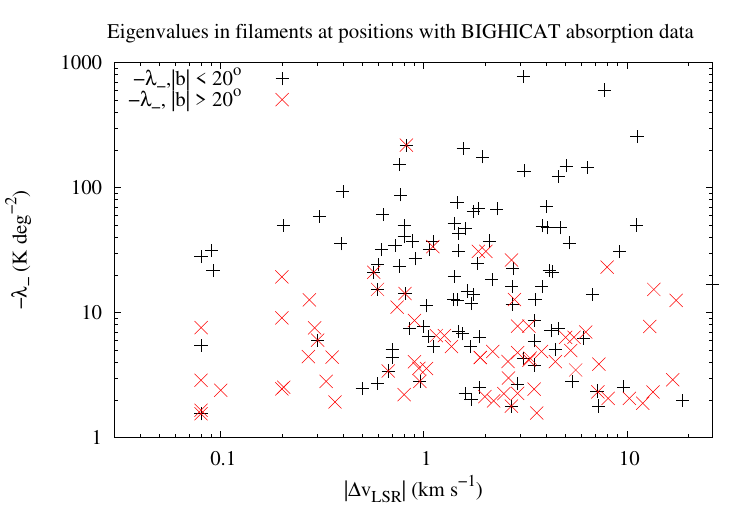}
   \caption{Parameters derived from BIGHICAT data for absorption
     components with closest match in filament velocities with
     deviations of $ |\Delta v_{\mathrm{LSR}}| $ \kms\ between
     absorption and emission components.
     Top: Spin temperatures $T_s$ in filaments. The horizontal line indicates
     the characteristic spin temperature $T_s = 50 $ K, derived by
     \citet{Heiles2005}. Bottom: Eigenvalues $-\lambda_-$ for the same
     sample.  }
   \label{Fig_Ts_lam}
\end{figure}

\begin{figure}[th] 
   \centering
   \includegraphics[width=9cm]{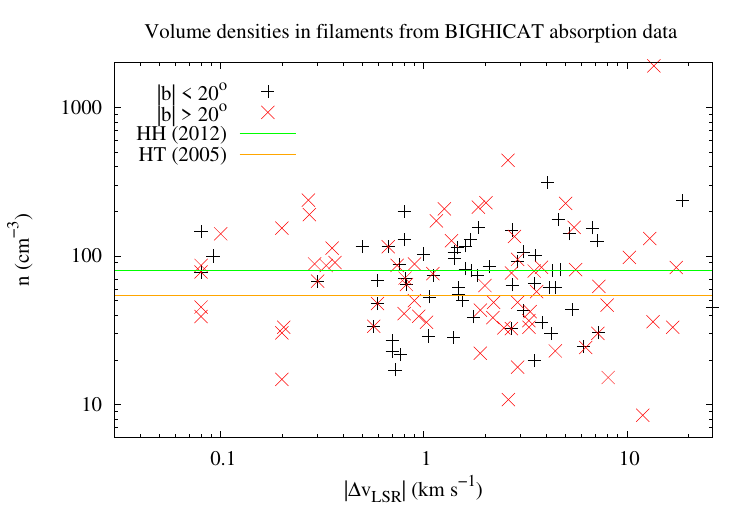}
   \caption{Volume densities $n$ in filaments from BIGHICAT data for
     absorption components with the closest match in filament velocities
     with deviations of $ |\Delta v_{\mathrm{LSR}}| $ \kms\ between
     absorption and emission components. The
     horizontal lines indicate the characteristic volume densities
     of $n = 80~{\rm cm^{-3}}$ derived by \citet{Heiles2012} and
     $n = 54~{\rm cm^{-3}}$ by \citet{Heiles2005}. }
   \label{Fig_nHI}
\end{figure}

\subsection{Spin temperatures and volume densities in \hi\ filaments }
\label{T_s}

In \citetalias{Kalberla2021} it was shown that filaments in \hi\ have
low Doppler temperatures. Such temperatures are derived from emission
data, and hence the observed line widths are broadened by turbulent motions
and the derived Doppler temperatures are only upper limits to the spin
temperatures of the \hi. With the advent of sensitive absorption surveys
it has become feasible to probe FIR/\hi\ filaments for spin
temperatures. \citet{Naomi2023} collected BIGHICAT, a catalog with 372
unique lines of sight, providing spin temperatures, optical depth, and
other properties. We use BIGHICAT to probe spin temperatures within
filaments and outside. Figure 3 of \citet{Naomi2023} shows the spatial
distribution of the absorption components.

BIGHICAT contains 66 positions without detectable absorption;  63 of these
positions (95\%) are located outside FIR/\hi\ filaments. These 63
positions probe only WNM, which  is known to be ubiquitous. The WNM is known
to contain no significant small-scale structures, and also has   a very low
optical depth. From the remaining sources 122 positions need to be
disregarded because the associated FIR eigenvalues exceed the threshold
$\lambda_- > -1.5\ \mathrm{K/deg}^{-2}$. A total of 37 positions from BIGHICAT
close to the Galactic equator are without available spin temperatures,
but in 23 of these cases we were able to replace these data by using
Gaussian components from the HI4PI survey. This way 174 positions from
the absorption survey could be used to determine absorption components
at velocities with the closest agreement to observed local filament
velocities. These components represent the local physical conditions
along the line of sight that are representative for significant
FIR/\hi\ filaments. Figure \ref{Fig_Ts_lam} shows the resulting spin
temperatures as a function of velocity deviation $ | \Delta
v_{\mathrm{LSR}}|$. These velocity deviations represent differences in
velocity between the pencil beam pointing to the background source and
the closest HI4PI survey position within a filament. Thus, the detectable
velocity differences $ | \Delta v_{\mathrm{LSR}}|$ represent turbulent
velocity fluctuations on scales below 3\farcm4.

The spin temperatures in Fig. \ref{Fig_Ts_lam} occupy a range $10 \la
T_s \la 500$ K, values that are, according to \citet{Wolfire2003},
characteristic for the temperature of the CNM. \citet{Heiles2005}
derived a typical CNM temperature of 50 K.  The lower display of
Fig. \ref{Fig_Ts_lam} shows that these low temperatures are associated
with significant FIR eigenvalues in $-\lambda_{-}$. There is a trend for
lower spin temperatures at positions with lower eigenvalues
$\lambda_{-}$, but the available data are currently insufficient to
establish a significant correlation (r = 0.33).
\citet{Naomi2023} expect that as soon as the Square Kilometer Array is
in full operation, the number of entries in BIGHICAT will increase by
several orders of magnitude. Thus, the significance of this correlation
can be tested within a few years. 

The typical conditions for the CNM in the local multiphase ISM in
thermal equilibrium with the WNM can, according to \citet{Jenkins2011}, be
determined from an equilibrium pressure of $\log (p/{\mathrm{k}}) =
3.58$. This results in a typical CNM exitation temperature of 80 K,
consistent with Fig. \ref{Fig_Ts_lam}. The cooling time for the ISM is
in this case estimated by \citet{Jenkins2011} to $3~10^4$ yr. This
timescale is short in comparison to the timescale for the growth of the
small-scale magnetic energy, characterized by the viscous eddy turnover
time in the order of $10^5$ yr \citep{Schekochihin2004}. The spin
temperature depends on excitation processes for the 21 cm line and can
for the CNM usually be equated to the exitation temperature
\citep[][Sect. 2]{Naomi2023}. Volume densities, derived from absorption
observations, are based on the assumption of a local thermal
equilibrium.  CNM filaments had sufficient time to reach thermal
equilibrium, and therefore BIGHICAT data can be used to derive realistic
volume densities. The results are shown in Fig. \ref{Fig_nHI}; the
volume densities are again typical for the CNM, and agree well with the
estimates $n = 54~{\rm cm^{-3}}$ and $n = 80~{\rm cm^{-3}}$ by
\citet{Heiles2005} or \citet{Heiles2012}, respectively. The available
sample of absorption features support the interpretation of filaments as
cold dense CNM structures. Outside the filaments no absorption and no
CNM is detectable.

\begin{figure}[ht] 
   \centering
   \includegraphics[width=9cm]{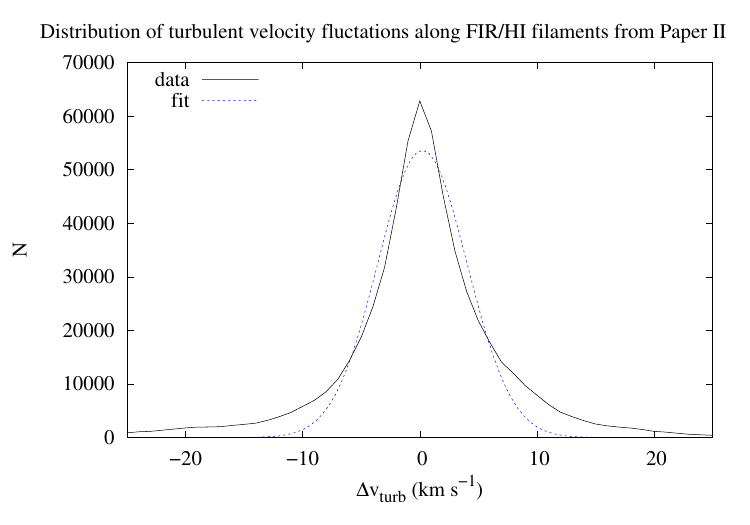}
   \caption{Distribution of turbulent velocity fluctuations $\Delta
     v_\mathrm{turb}$ along filaments for the sample used in
     \citetalias{Kalberla2023}. The Gaussian fit has a velocity
     dispersion of $\sigma = 3.8 \pm 0.1 $ \kms.}
   \label{fig_dvel}
\end{figure}

\begin{figure*}[th] 
   \centering
   \includegraphics[width=6cm]{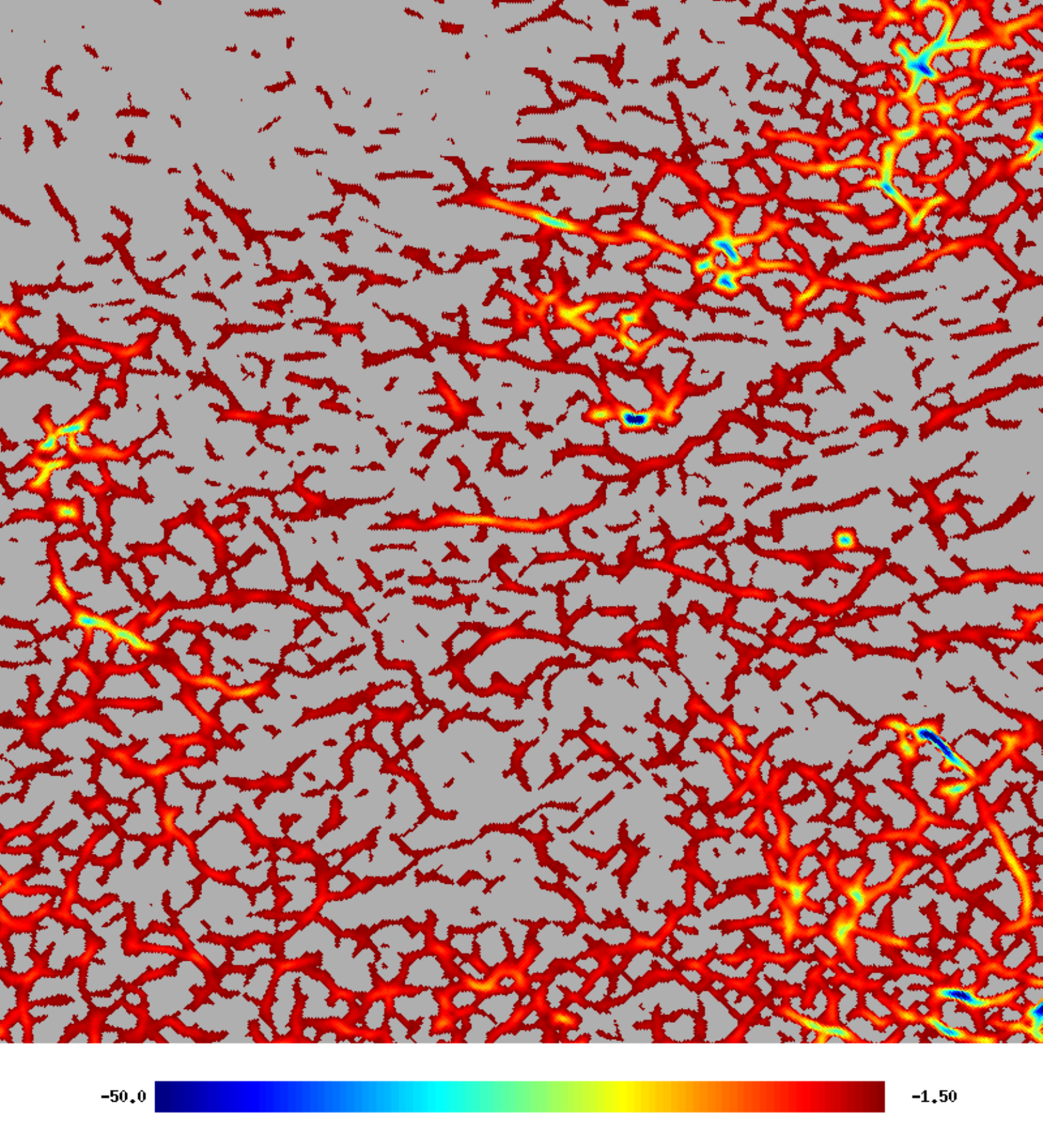}
   \includegraphics[width=6cm]{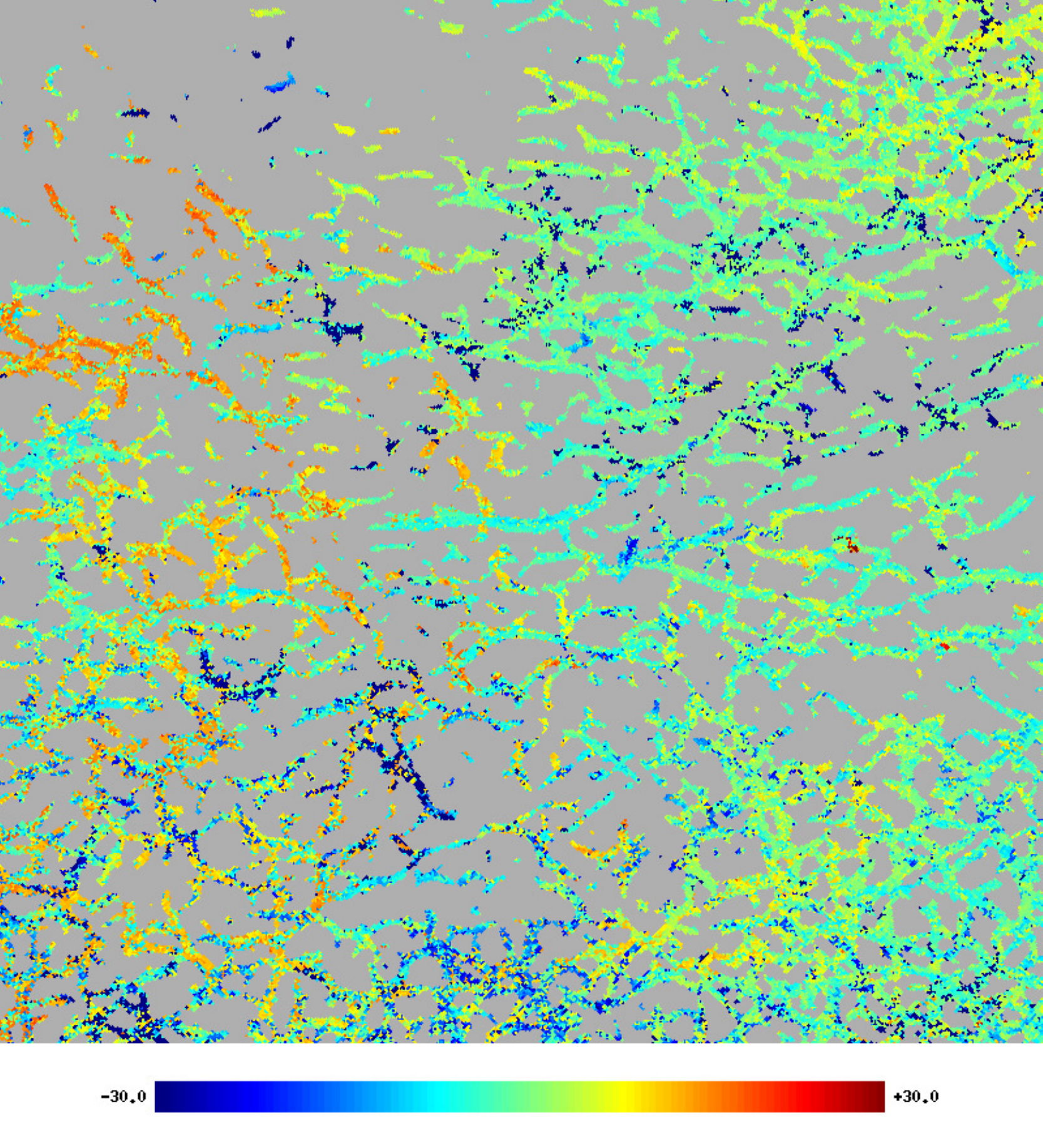}
   \includegraphics[width=6cm]{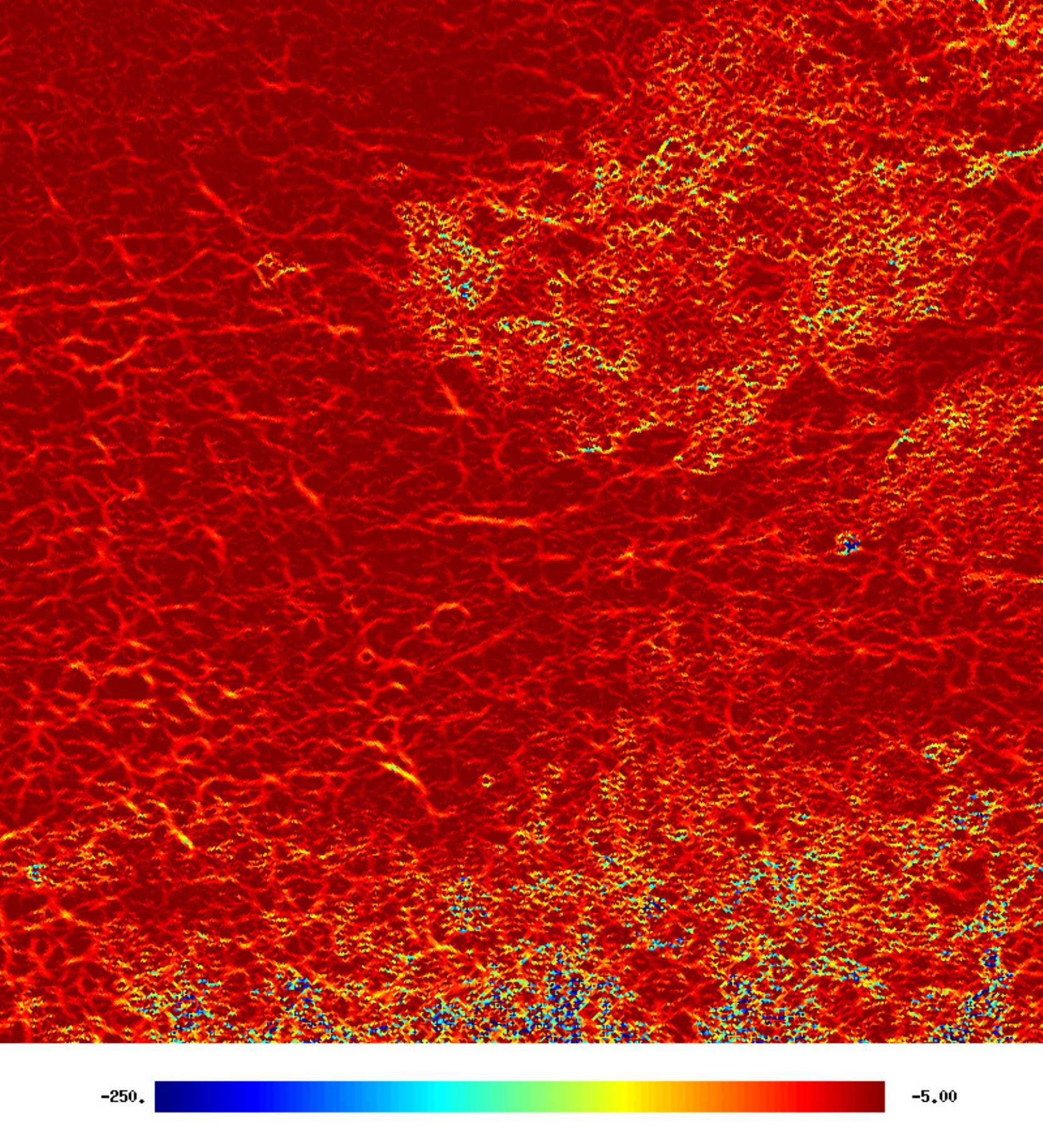}
   \caption{ Examples for structures derived with different methods,
     coordinates are identical with Fig. \ref{Fig_NoiseMaps_VDA}.  Left:
     filaments derived for 857 GHz FIR data in
     \citetalias{Kalberla2023}, shown are eigenvalues $-50 < \lambda_- <
     -1.5\ \mathrm{K/deg}^{-2}$. Center: PPV$_\mathrm{fil}$ with color
     coding in velocity, representing the corresponding velocity field
     in the range $ -30 < v_{\mathrm{LSR}} < 30 $ \kms. Right: caustics
     in \hi\ column density maps from \citetalias{Kalberla2023},
     integrated from $ -50 < v_{\mathrm{LSR}} < 50 $ \kms\ with
     eigenvalues $-250 < \lambda_- < -5\ \mathrm{K/deg}^{-2}$. }
   \label{Fig_857_HI}
\end{figure*}

\subsection{Turbulent velocity fluctuations }
\label{v_turb}

As detailed in Sect. \ref{Hessian}, \hi\ filaments were matched to FIR
filaments using both Hessian eigenvalues and orientation angles
(deduced from eigenvectors). In \citetalias{Kalberla2023} the resulting
set of Morse complexes was decomposed in 6568 individual
FIR/\hi\ filaments. For each of these filaments an average radial
velocity $\langle v_\mathrm{fil} \rangle$ was determined.  Along each of
the \hi\ filaments fluctuating radial velocities were observed.  We
consider these fluctuations as an imprint of the turbulent velocity
field $\Delta v_\mathrm{turb}(l,b) = ( v_\mathrm{fil}(l,b) - \langle
v_\mathrm{fil} \rangle)$ by subtracting for each filament position the
average radial velocity $\langle v_\mathrm{fil} \rangle $ from the local
filament velocity $v_\mathrm{fil}(l,b)$. Figure \ref{fig_dvel} shows the
distribution of $\Delta v_\mathrm{turb}$. The observed turbulent
velocity dispersion is $\sigma = 3.8 \pm 0.1 $ \kms.  This is low in
comparison to the expected dispersion $\sigma \sim 10 $ \kms\ in the
turbulent ISM \citep{Burkert2006}. The observed $\Delta v_\mathrm{turb}$
distribution is  not uniform, however; the median dispersion is $\Delta v
= 5.24$ \kms. Figure 12 in \citetalias{Kalberla2023} shows that
filaments with low aspect ratios can have a low velocity dispersion but
prominent filaments, with high aspect ratios and covering large surface
areas, have enhanced velocity dispersions. These approach   the limit
$\sigma \sim 10 $ \kms. We also note the extended wings in
Fig. \ref{fig_dvel}, indicating some local misidentifications in the
FIR/\hi\ filaments causing uncertainties in the determination of
filament velocities.

Extended cold \hi\ filaments with a typical linewidth of 3
\kms\ (e.g., \citetalias{Kalberla2023}, Sect. 3.1) that are exposed to
turbulent velocity fluctuations of 3.8 to 10 \kms\ are affected necessarily
by shifts in observed radial velocities. Filaments that are observed
in channel maps with a velocity width of 1 \kms\ appear disrupted,
and filaments appear to have low aspect ratios. Within PPV channel maps,
filament elements that are affected by turbulent velocity fluctuations
shift frequently from one velocity channel to another. This leads to an
apparent fragmentation and lowers aspect ratios if only channel maps are
considered.

Tracing the FIR orientation angle in \hi\ recovers the
PPV$_\mathrm{fil}$ filament geometry in the plane of the sky. Both FIR
and \hi\ filaments share common orientation angles. FIR filaments are
observed in broadband and the FIR orientation angles are not affected by
velocity fluctuations. Tracing FIR orientation angles in \hi\ by fitting
(velocity dependent) orientation angles tells us the velocity of the FIR
filaments (\citealt{Clark2019b} and \citetalias{Kalberla2023},
Sect. 3.3). Thus, the 3D PPV geometry is modified for each individual
filament to the 2D case PPV$_\mathrm{fil}$, where V$_\mathrm{fil}$
stands for the position-dependent filament velocity across the PPV
cube. The transformation from PPV to PPV$_\mathrm{fil}$ represents for
each individual filament a smooth transformation (or deformation) of the
velocity space.  Magnetic fields  associated with the filaments suffer
from foldings that cause magnetic tension forces with back reactions on
the turbulent flow \citep{Schekochihin2004}. In such a scenario, density
structures need to be understood as structures in
$l,b,v_{\mathrm{LSR}}$, and hence with PPV$_\mathrm{fil}$ morphology. Each
filament occupies an individual PPV$_\mathrm{fil}$ space. The width of
the V$_\mathrm{fil}$ slice depends on the observed CNM line width, and is
typically 3 \kms. The Morse complexes that we analyzed   need to be
understood as manifolds in PPV$_\mathrm{fil}$. Since the filaments are
disjunct in position, an all-sky PPV$_\mathrm{fil}$ distribution can be
derived.

\subsection{VCA velocity mapping in slices}
\label{mapping}

VCA and VDA are theories that are strictly designed for a PPV geometry,
decomposing the emission in velocity slices of variable thickness. The
claim is that VCA allows a separation of velocity and density fields
that affect the PPV data cube, depending on the thickness of the
velocity slice. This theory does not allow a local transition of
distinct homogeneous filament structures from PPV to
PPV$_\mathrm{fil}$. The expected role of (hydrodynamic) turbulence is to
induce velocity crowding. Filamentary structures are understood as
velocity caustics, caused by velocity crowding along the line of
sight. Spatially distributed \hi\ along the line of sight is affected by
turbulent velocities and mimics  density structures in this way. Thus,
turbulence moves separate \hi\ components into the same velocity
coordinate in PPV space, leading to a merging of these components into a
single structure (e.g., \citealt[][Fig. 1]{Hu2023}).  VCA velocity mapping
does not allow us to interpret filamentary structures as real physical
objects.

The spectrum of eddies that correspond to most of the turbulent energy
at large scales corresponds to the spectrum of thin channel map
intensity fluctuations having most of the energy at small scales
\citep{Kandel2016}.  \citet{Lazarian2018} conclude that on scales larger
than 3 pc, the intensity of fluctuations in the channel maps is produced
to a significant degree by velocity caustics rather than real physical
entities (i.e., filaments). This 3 pc scale, derived by
\citet{Lazarian2018} for a turbulent energy cascade with a steep power
spectrum, can be compared with the filament widths of 0.63 pc derived in
\citetalias{Kalberla2023} for an average filament distance of 250
pc. Velocity mapping is affected by projection effects, but must be
spatially coherent in direction perpendicular to the filament. HI4PI
data constrain this coherence to 0.63 pc, otherwise the filamentary
structures would be smoothed out.

The VCA velocity mapping of distributed \hi\ components along the line
of sight to the observed PPV space has consequences for models of the
internal structure of FIR filaments.  Broadband observations of thermal
dust emission are sensitive to the density field, but not to
velocity. For FIR/\hi\ filaments the velocity crowding effect implies
that several distinct turbulence induced dust components must exist
along the line of sight. This is in clear contrast to FIR/\hi\ filaments
as coherent structures in PPV$_\mathrm{fil}$. \citet{Planck2016} consider
FIR filaments in the diffuse ISM at high Galactic latitudes as density
structures;  in addition, \citet{Planck2016a} consider them as ridges that are at high Galactic
latitudes usually aligned with the magnetic field measured on the
structures.

\begin{figure*}[ht] 
   \centering
   \includegraphics[width=6cm]{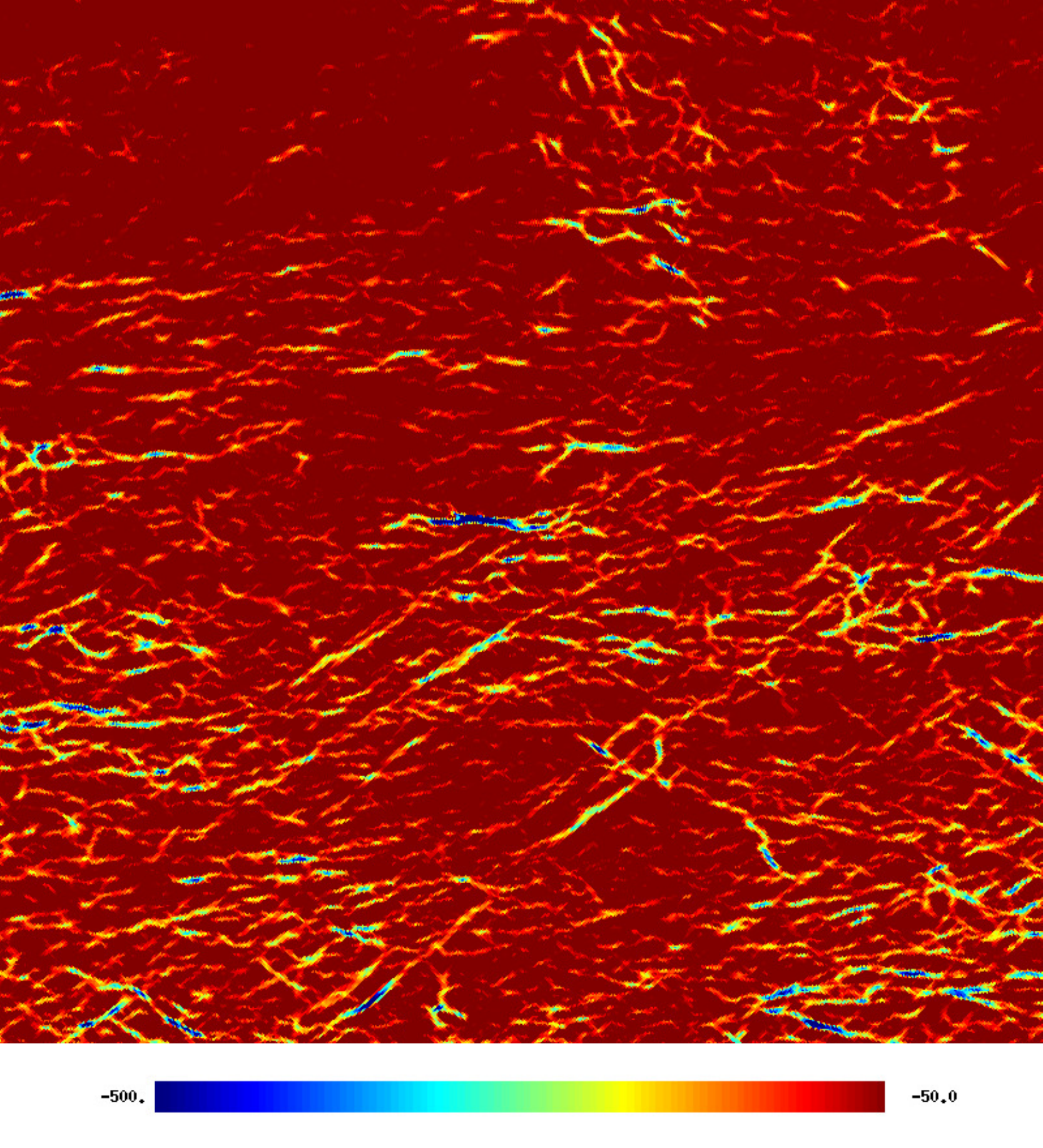}
   \includegraphics[width=6cm]{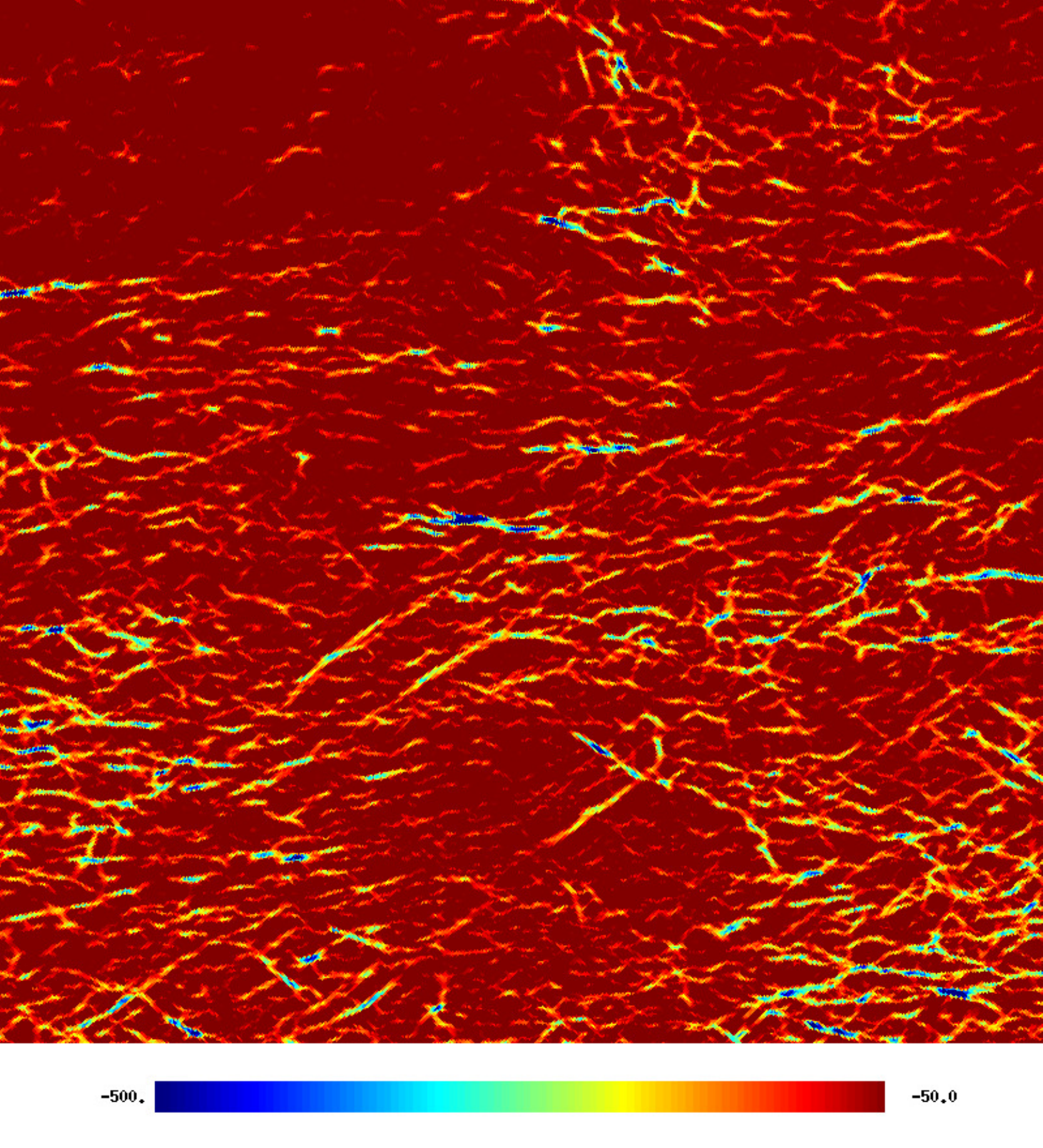}
   \includegraphics[width=6cm]{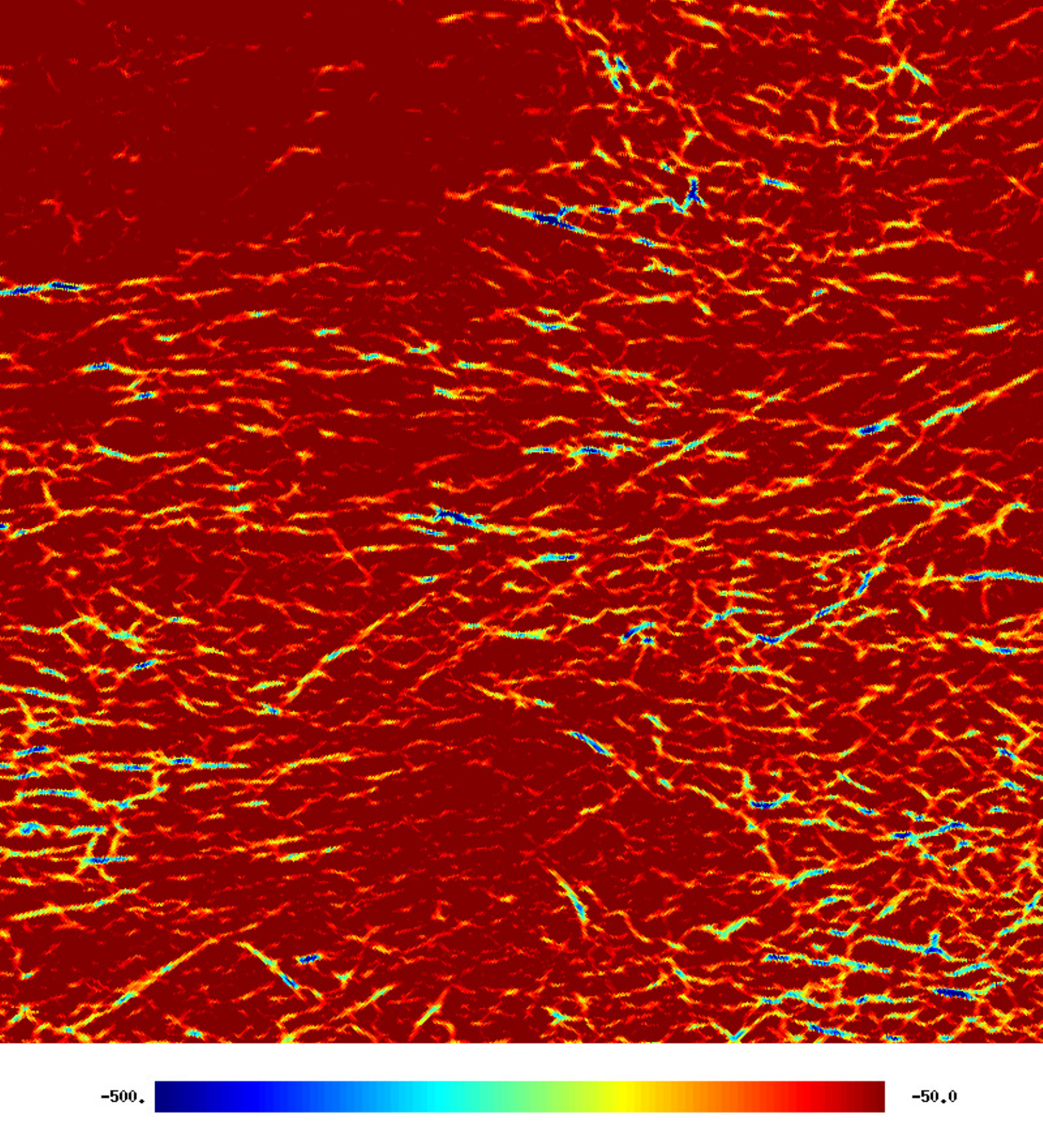}
   \includegraphics[width=6cm]{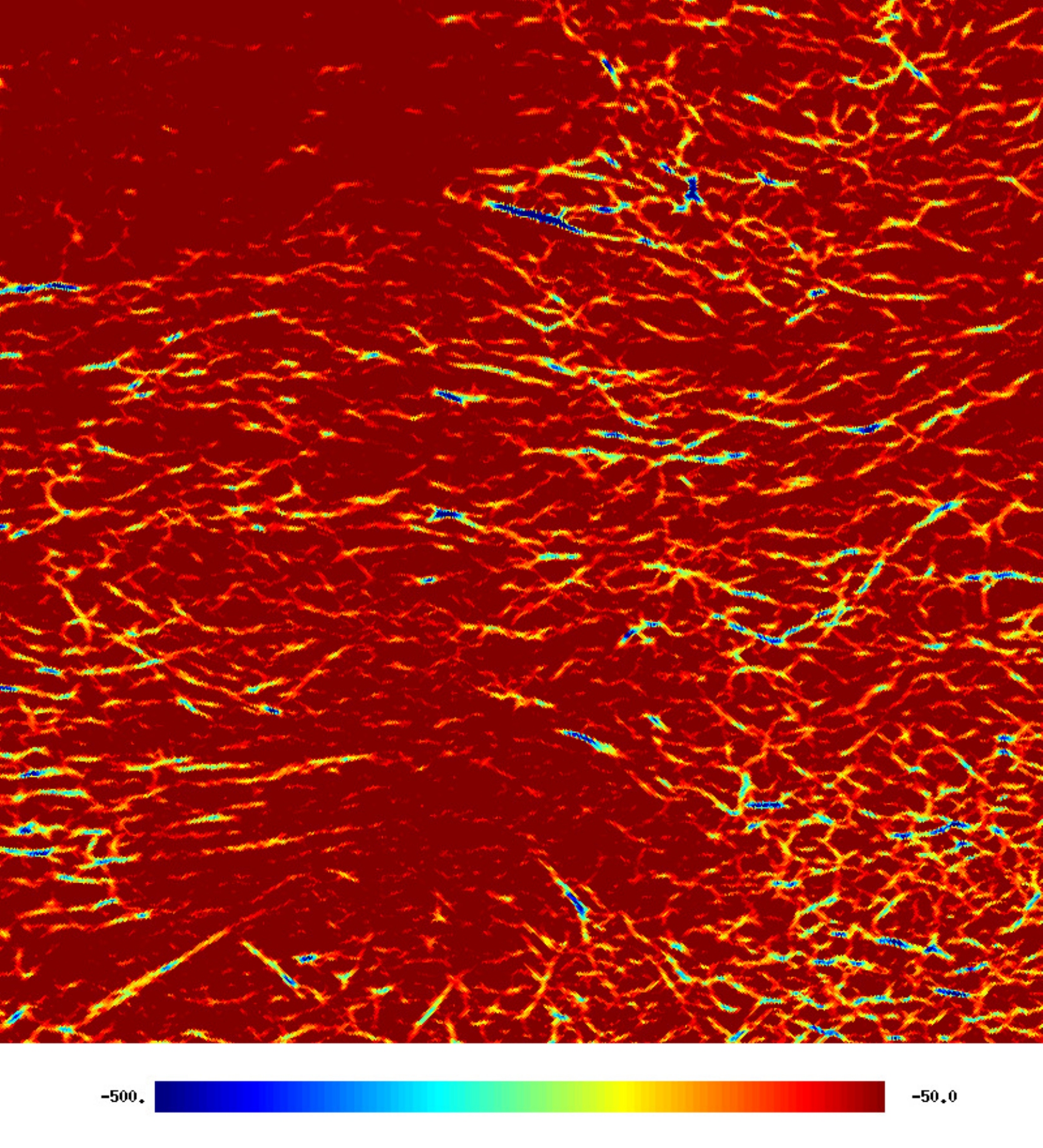}
   \includegraphics[width=6cm]{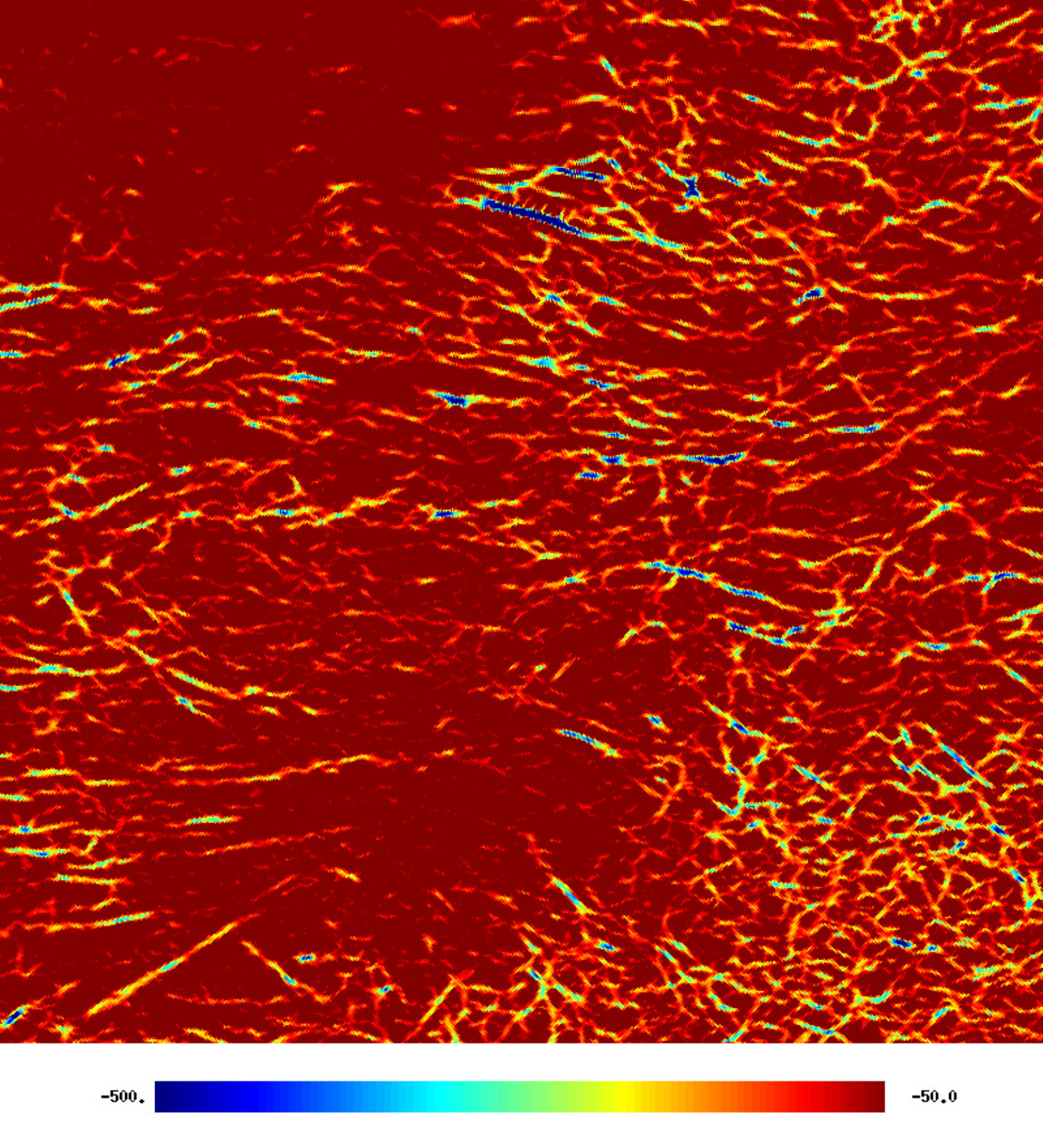}
   \includegraphics[width=6cm]{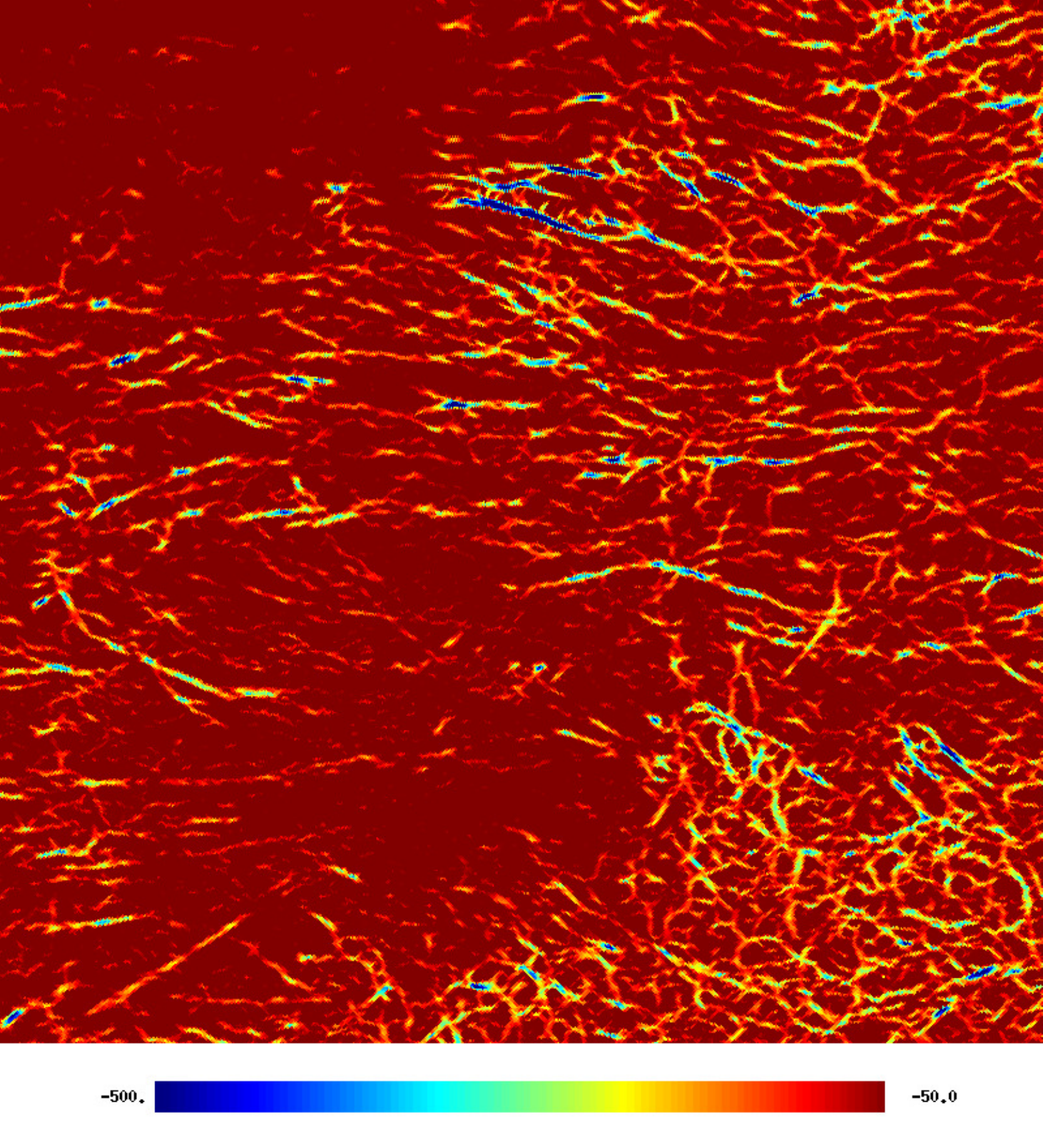}
   \includegraphics[width=6cm]{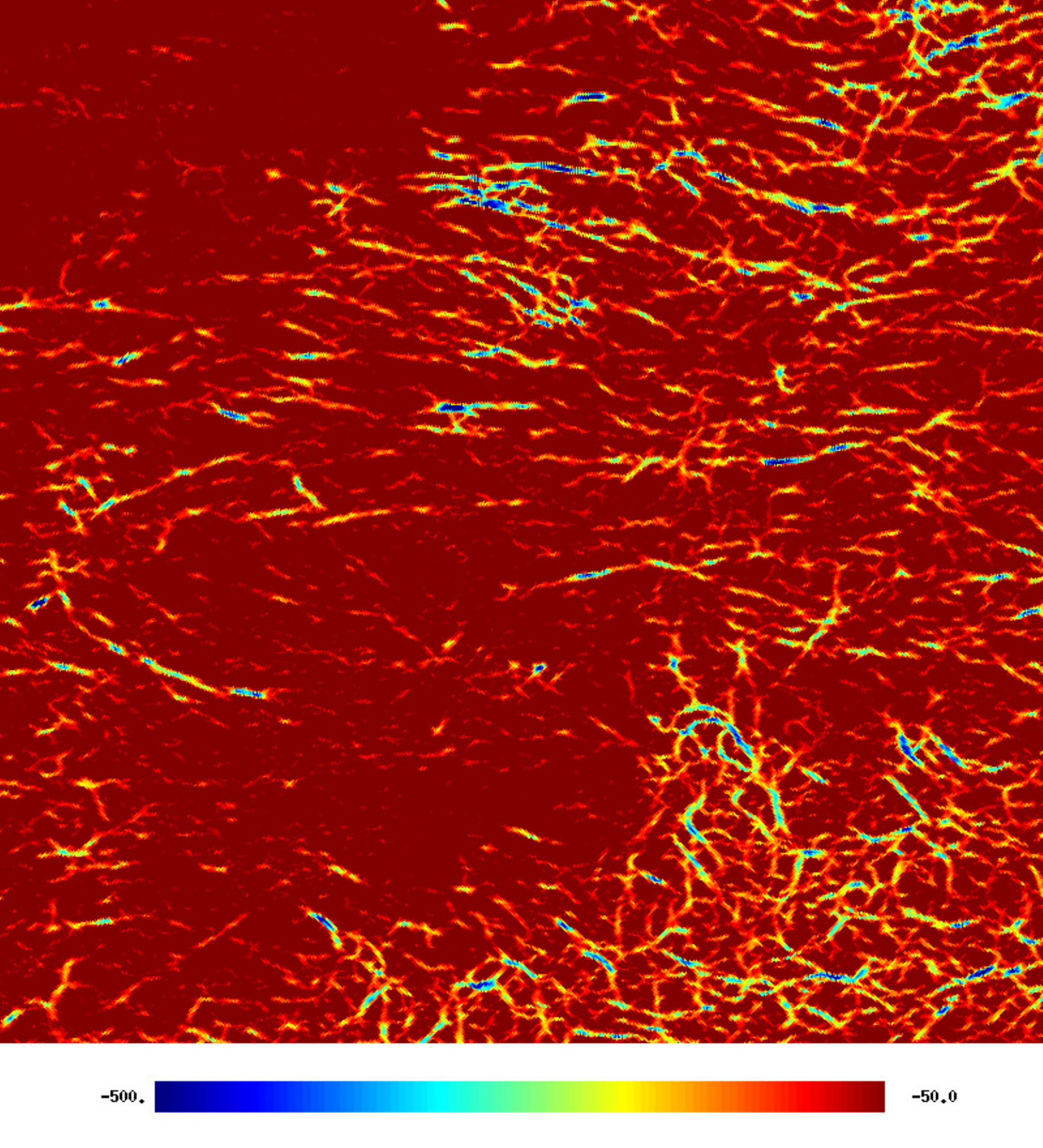}
   \includegraphics[width=6cm]{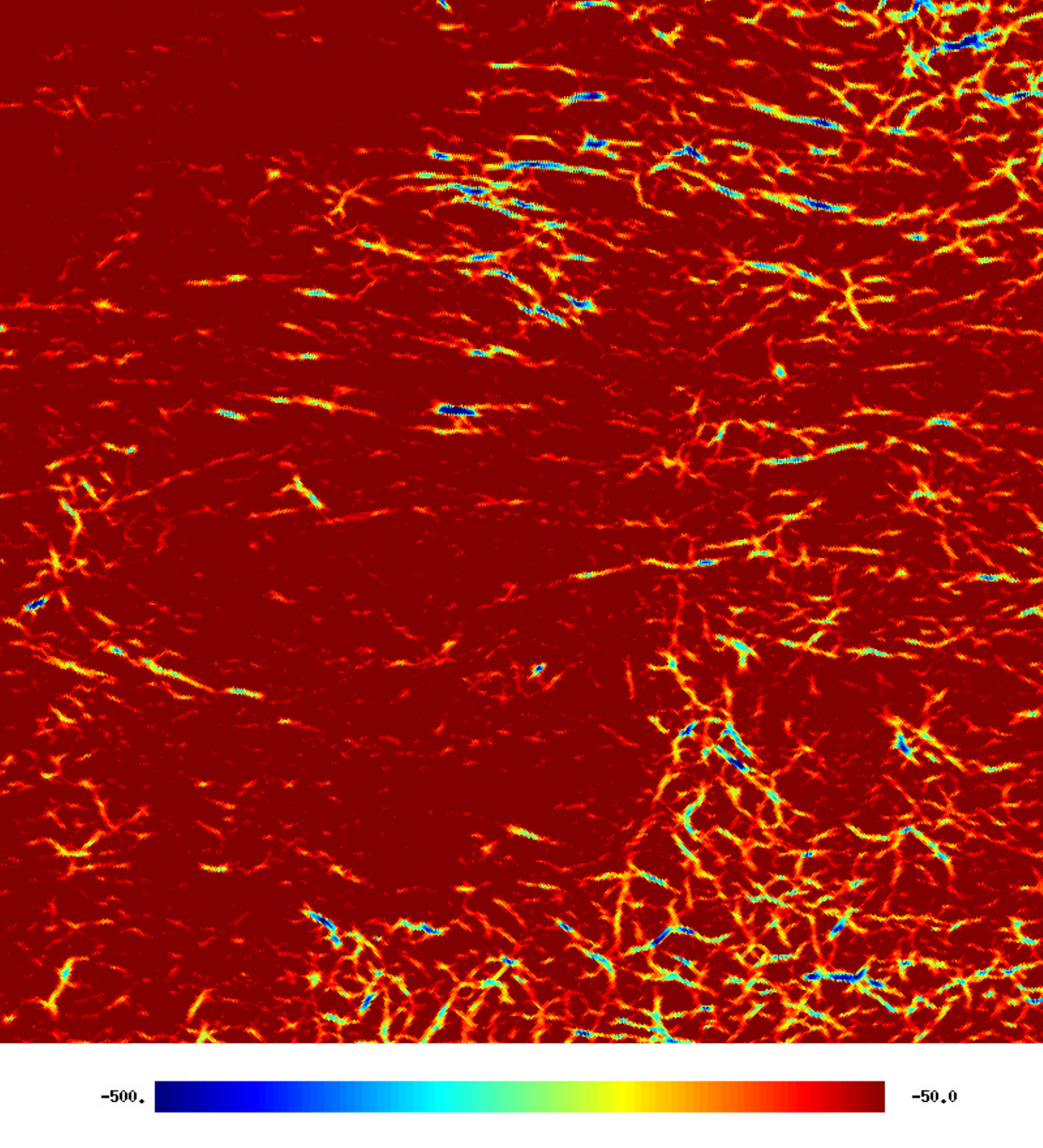}
   \includegraphics[width=6cm]{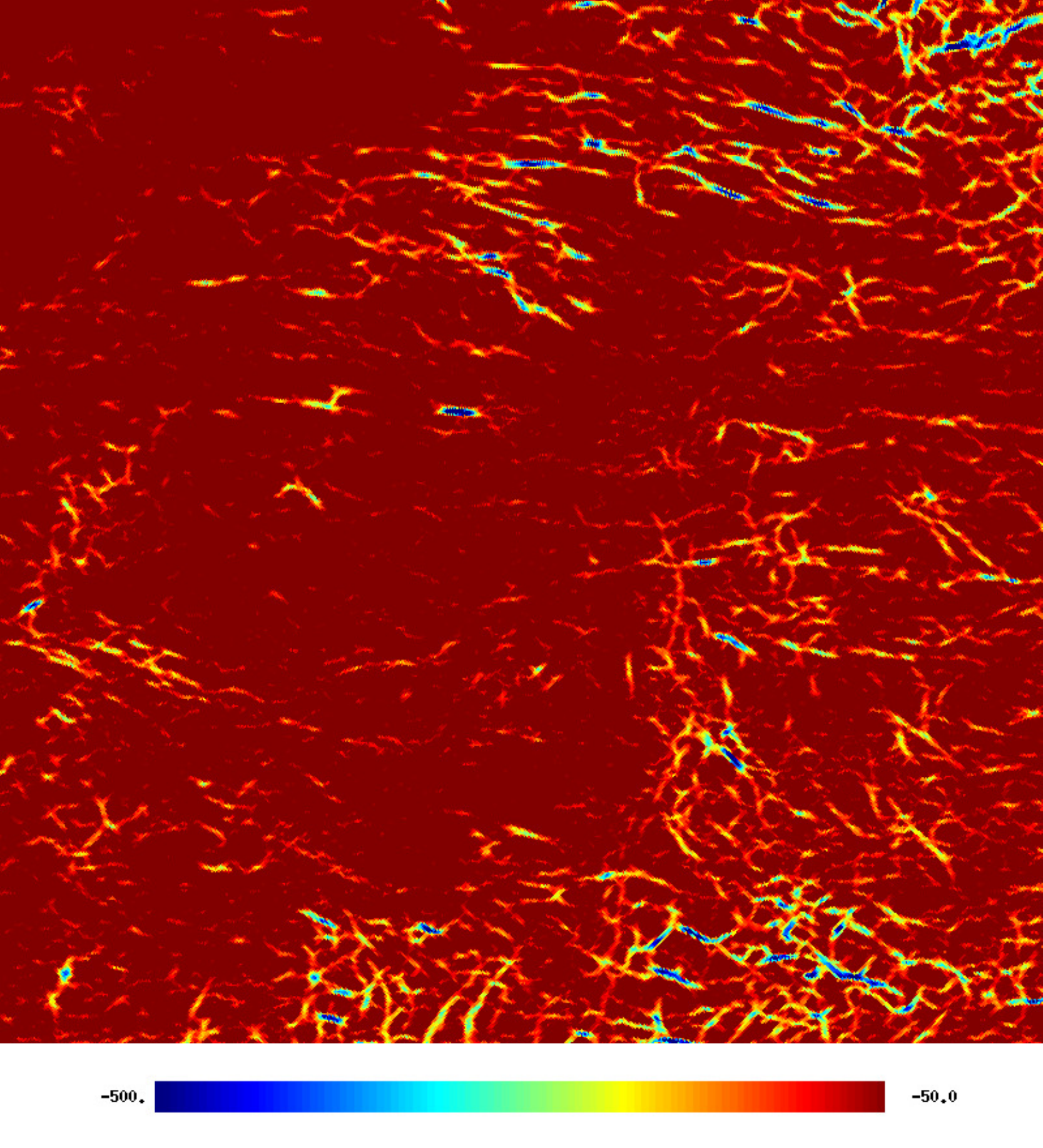}
   \caption{PPV maps of filamentary structures at $l = 160\degr$, $b =
     30\degr$ in gnomonic projection. The coordinates are identical to those in      Fig. \ref{Fig_NoiseMaps_VDA}. Shown are \hi\ eigenvalues
     $\lambda_{-}$ for $p$ in channel maps with a 1 \kms\ channel width
     at velocities from -4 \kms\ (top left) to 4 \kms\ (bottom right).
   }
   \label{Fig_HI}
\end{figure*}

\begin{figure*}[ht] 
   \centering
   \includegraphics[width=6cm]{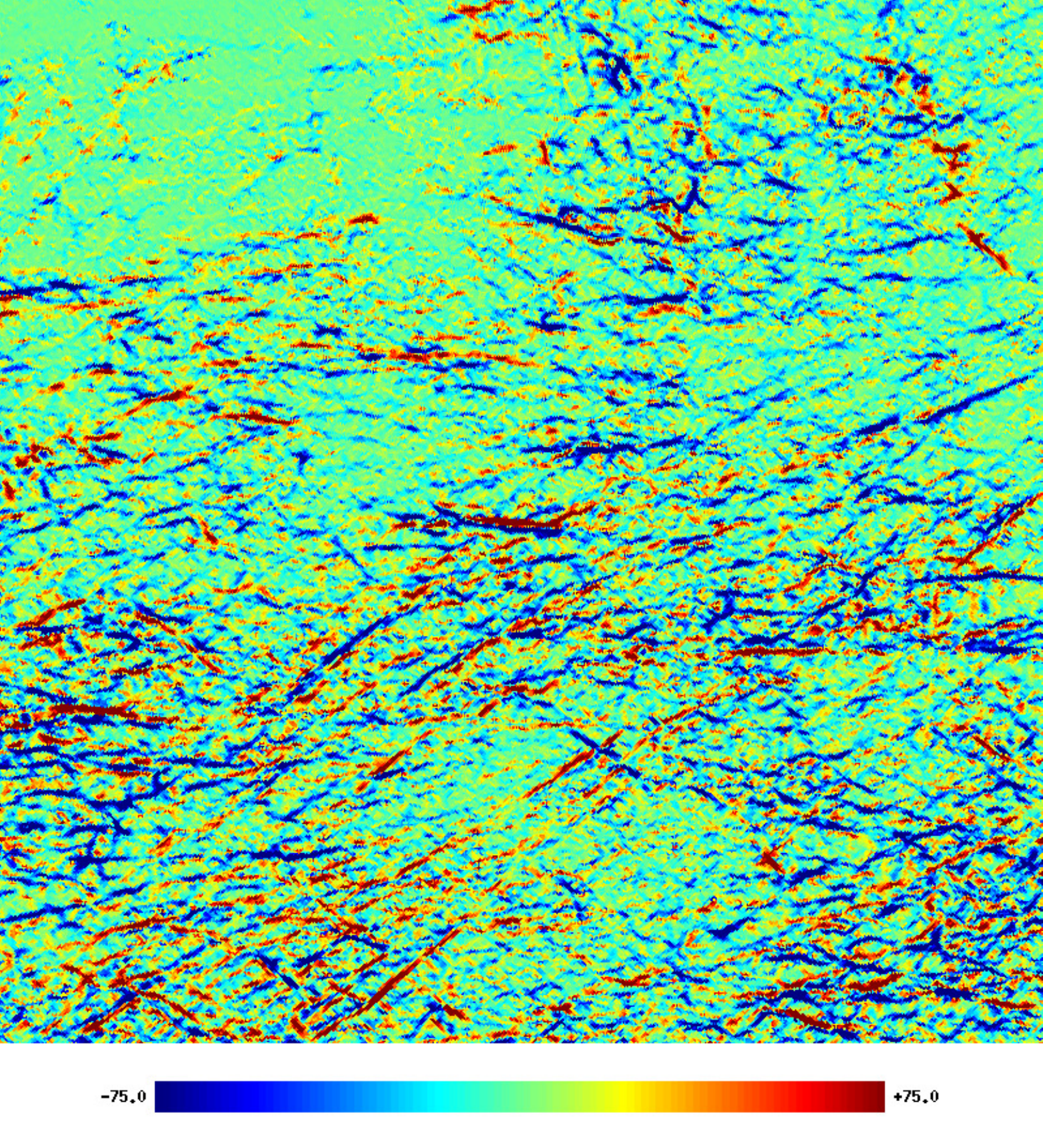}
   \includegraphics[width=6cm]{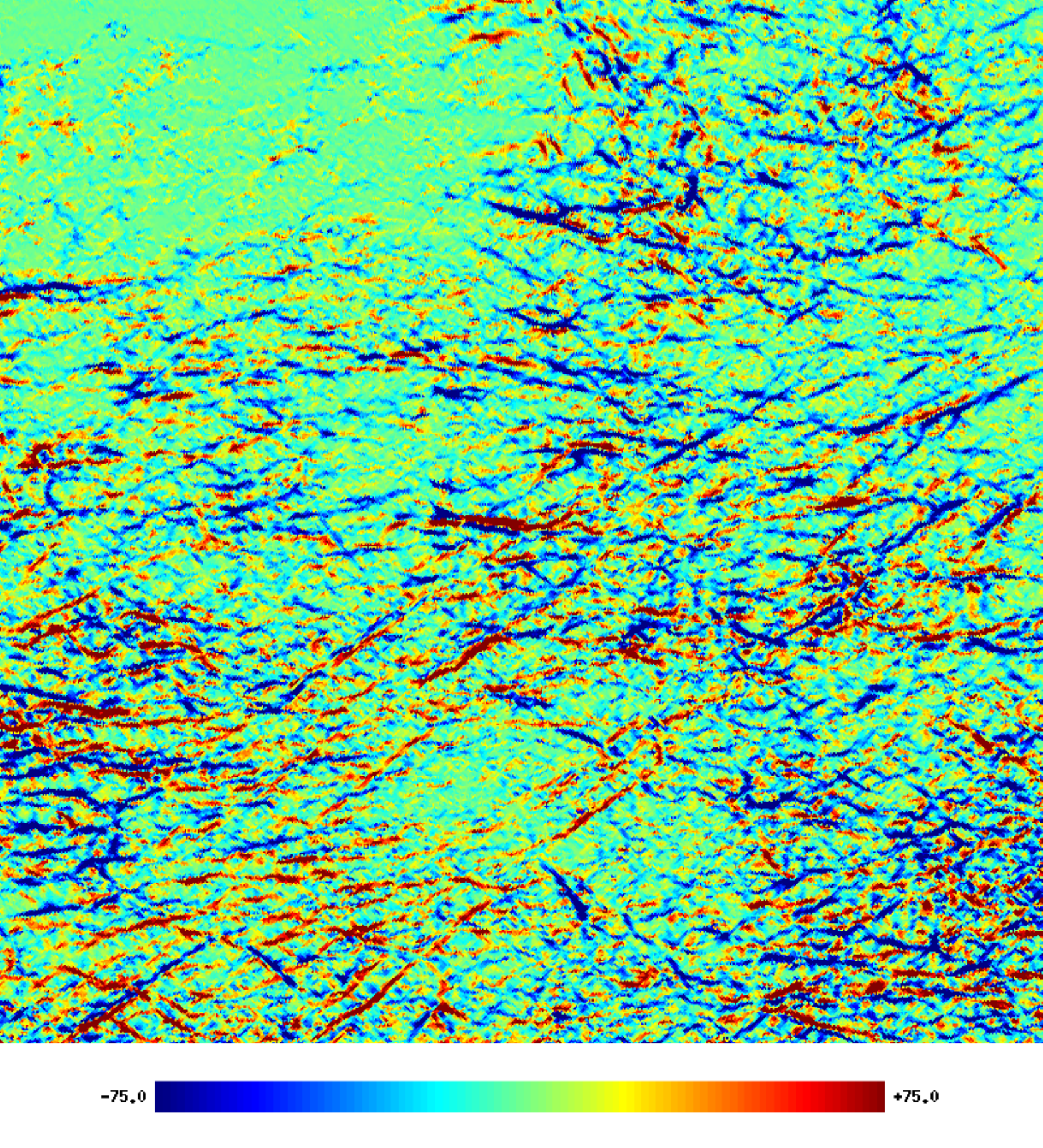}
   \includegraphics[width=6cm]{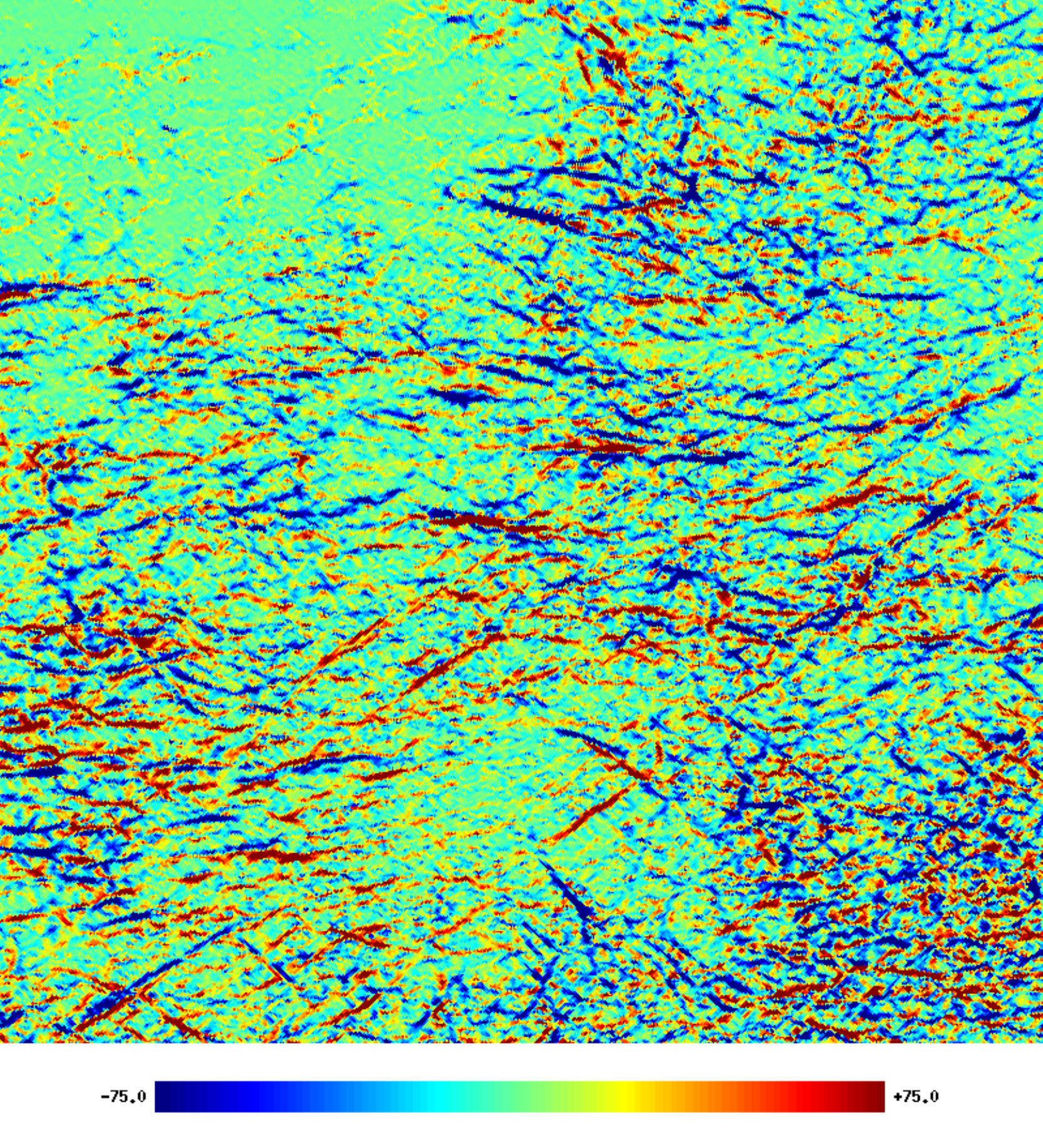}
   \includegraphics[width=6cm]{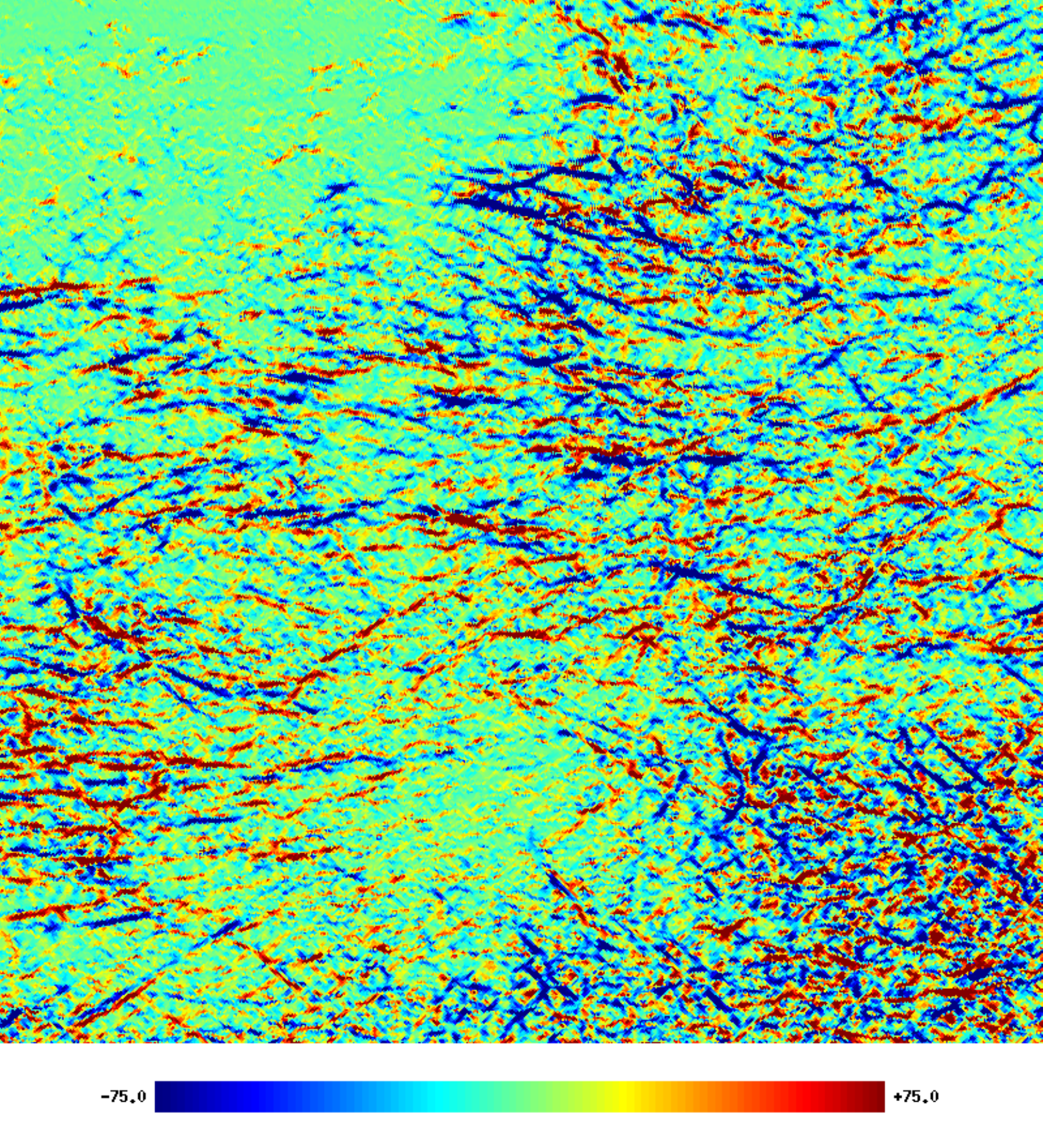}
   \includegraphics[width=6cm]{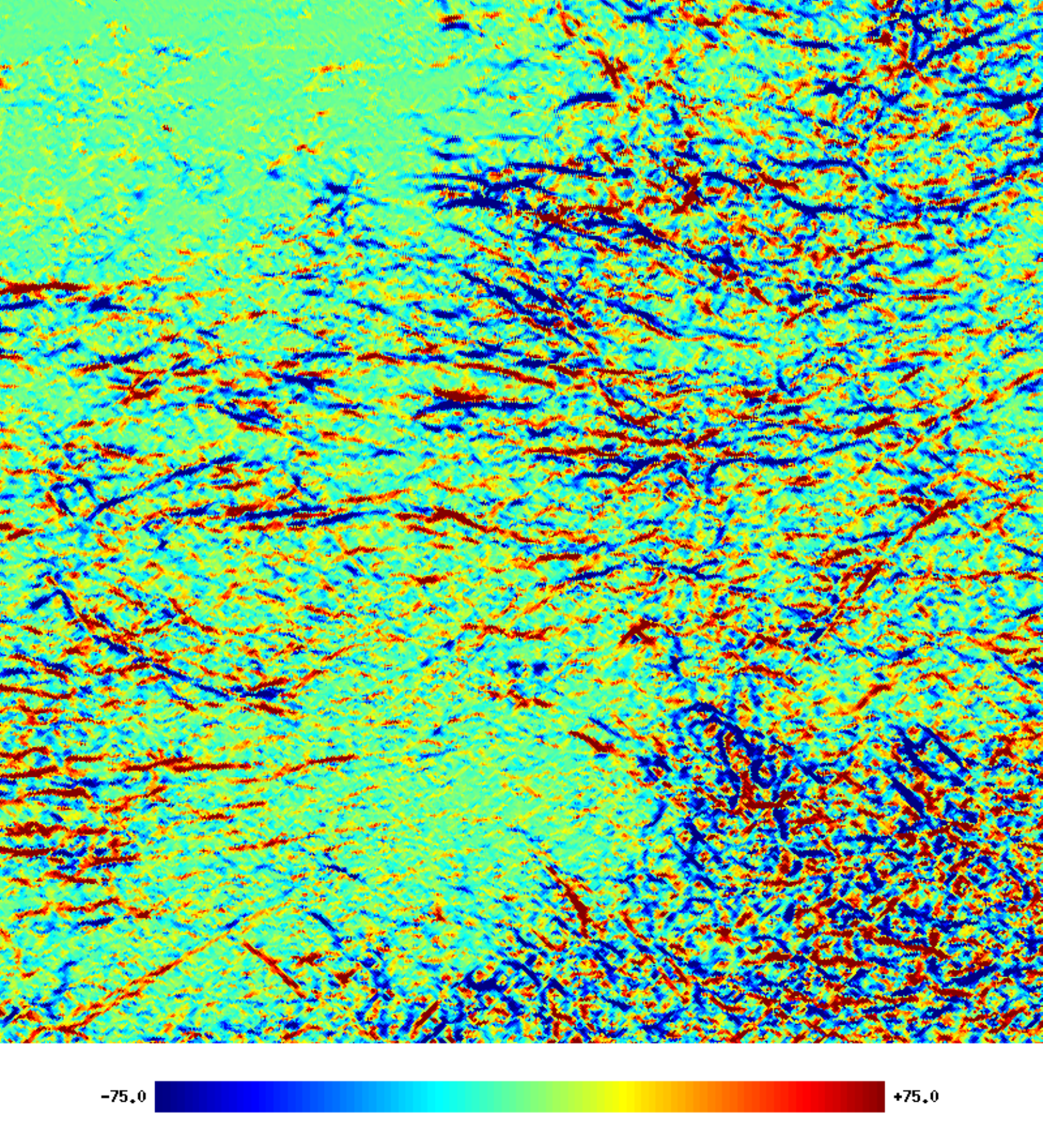}
   \includegraphics[width=6cm]{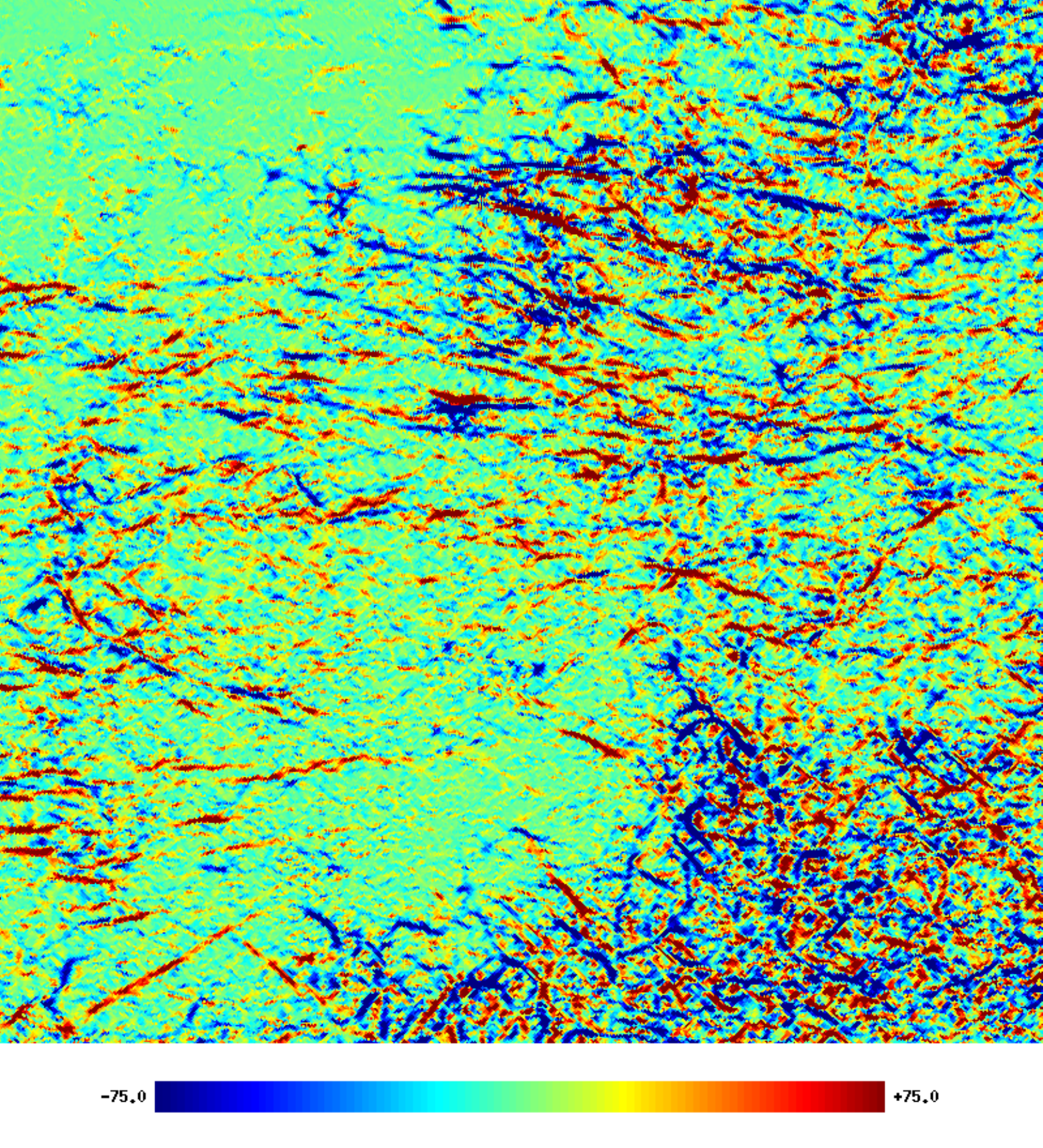}
   \includegraphics[width=6cm]{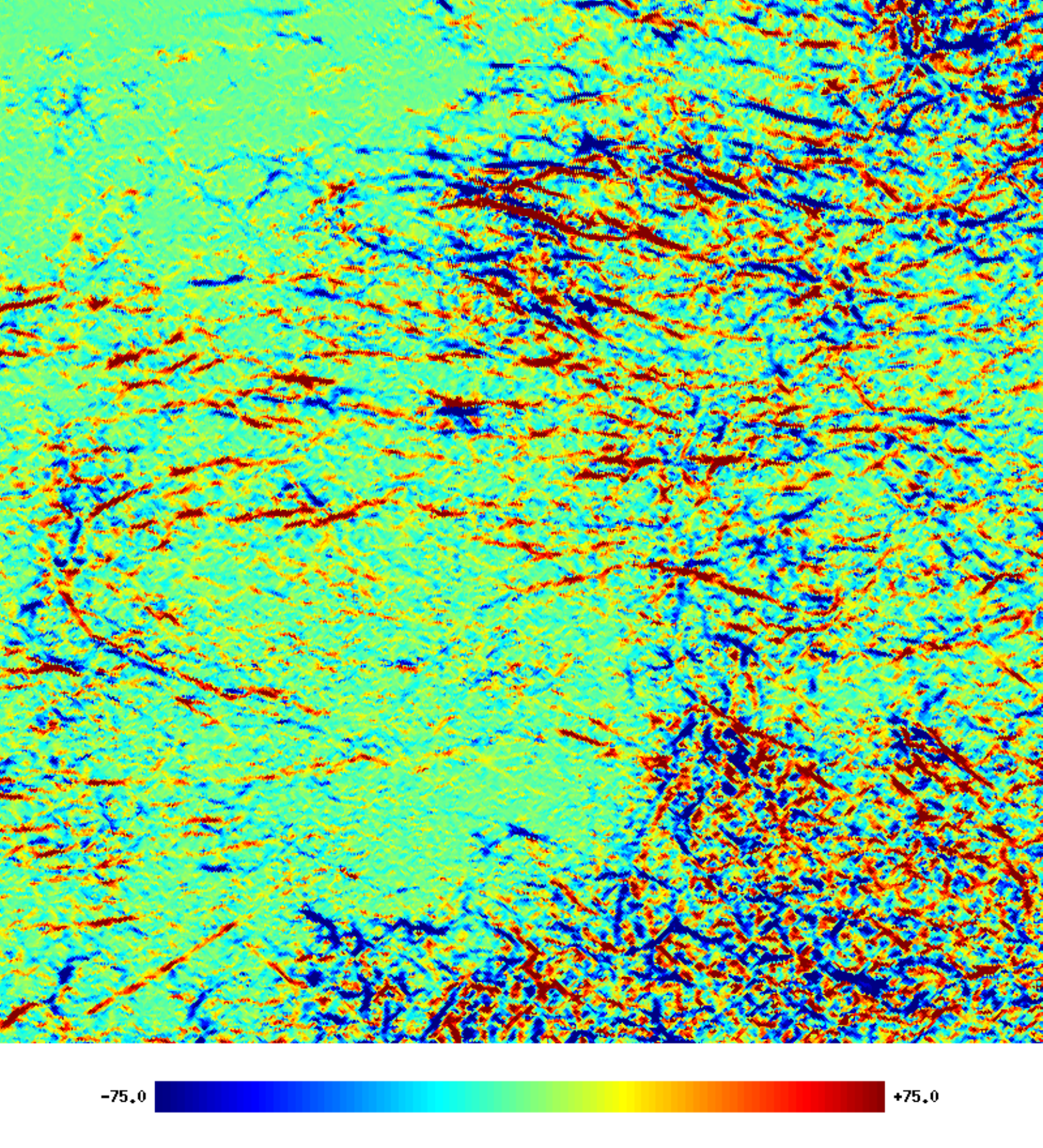}
   \includegraphics[width=6cm]{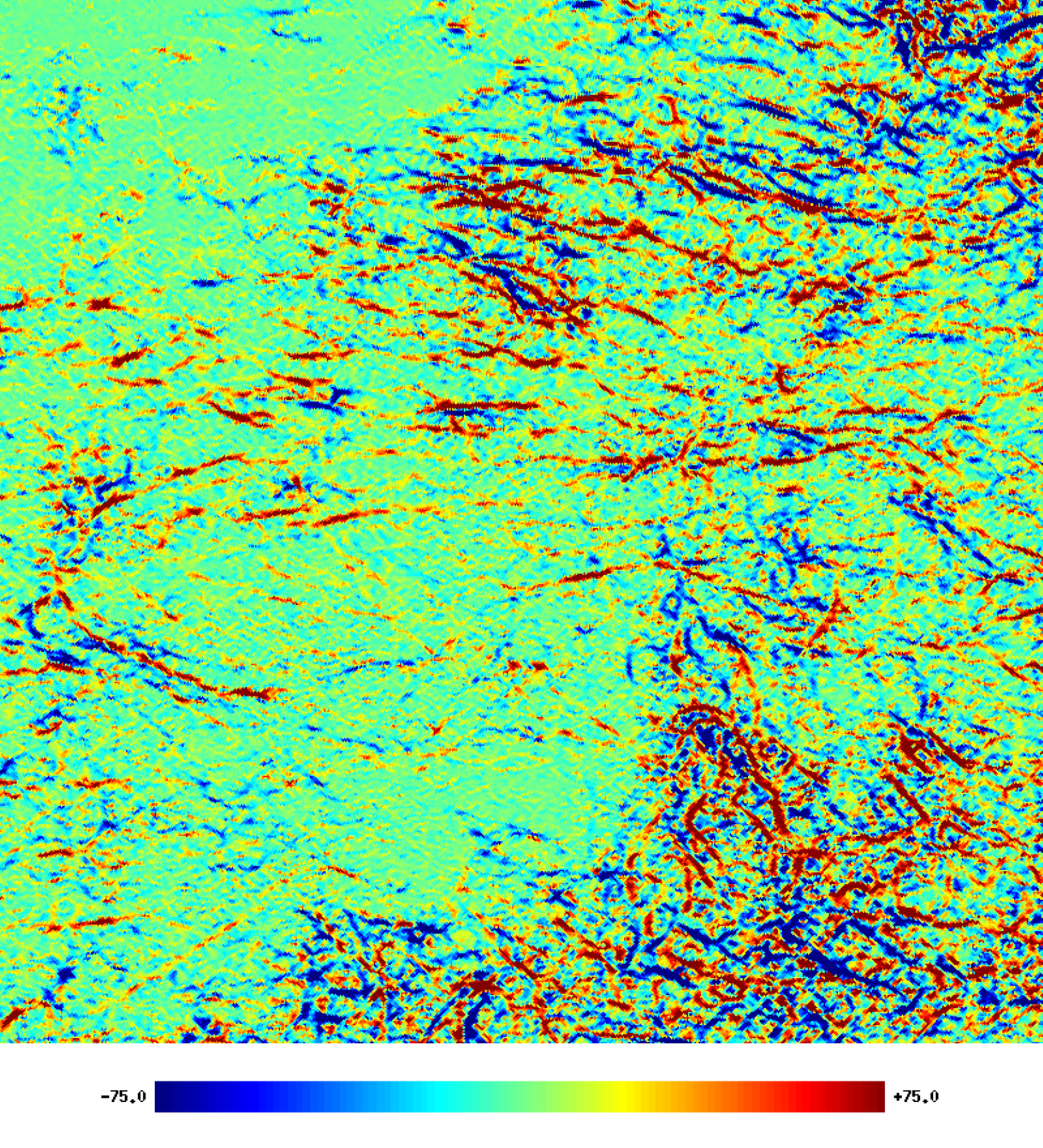}
   \includegraphics[width=6cm]{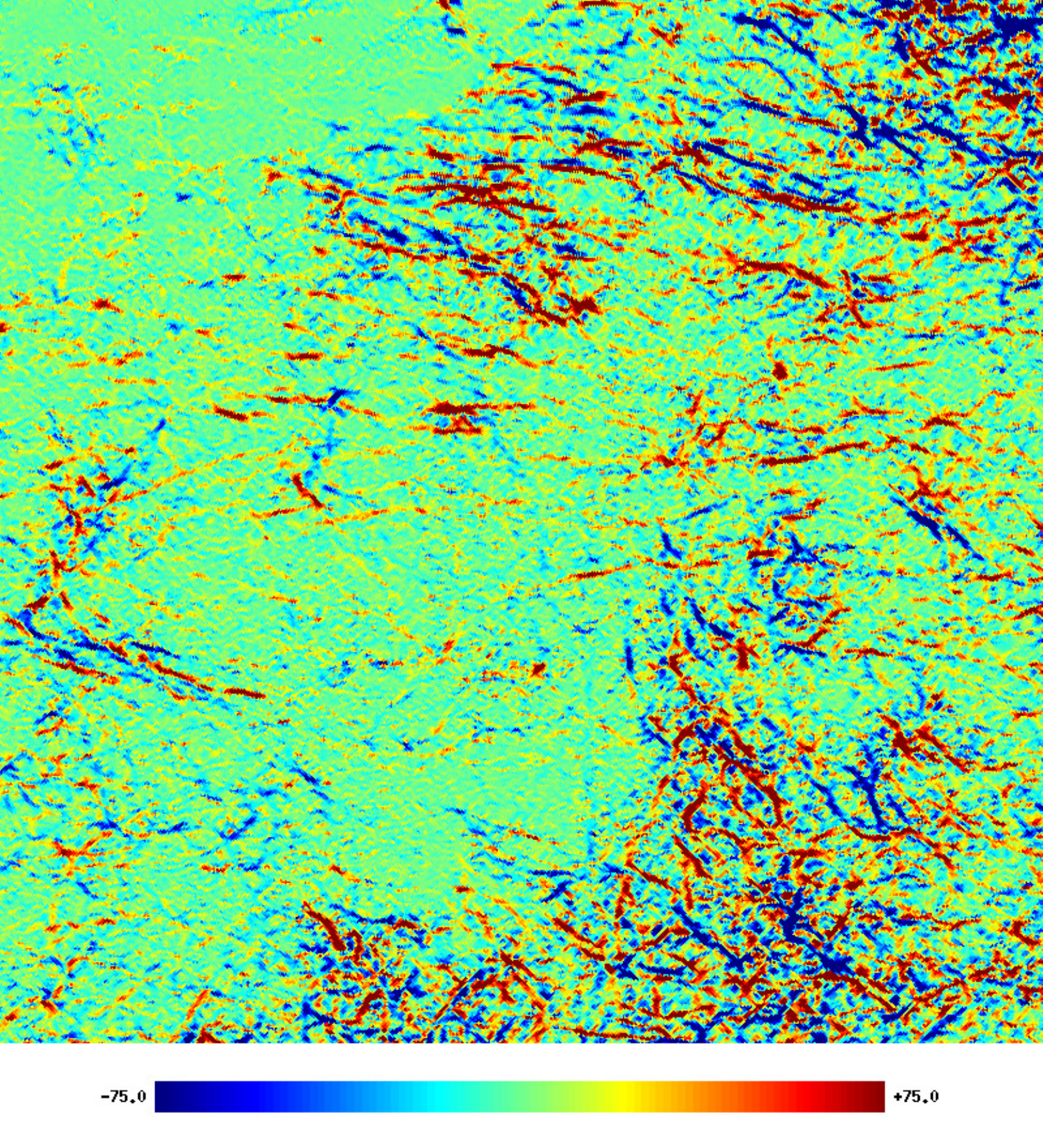}
   \caption{PPV velocity gradient maps according to Eq. (\ref{eq:VGT})
     for the same velocity channels   displayed in Fig. \ref{Fig_HI}.
   }
   \label{Fig_HI:diff}
\end{figure*}

\begin{figure}[th] 
   \centering
   \includegraphics[width=9cm]{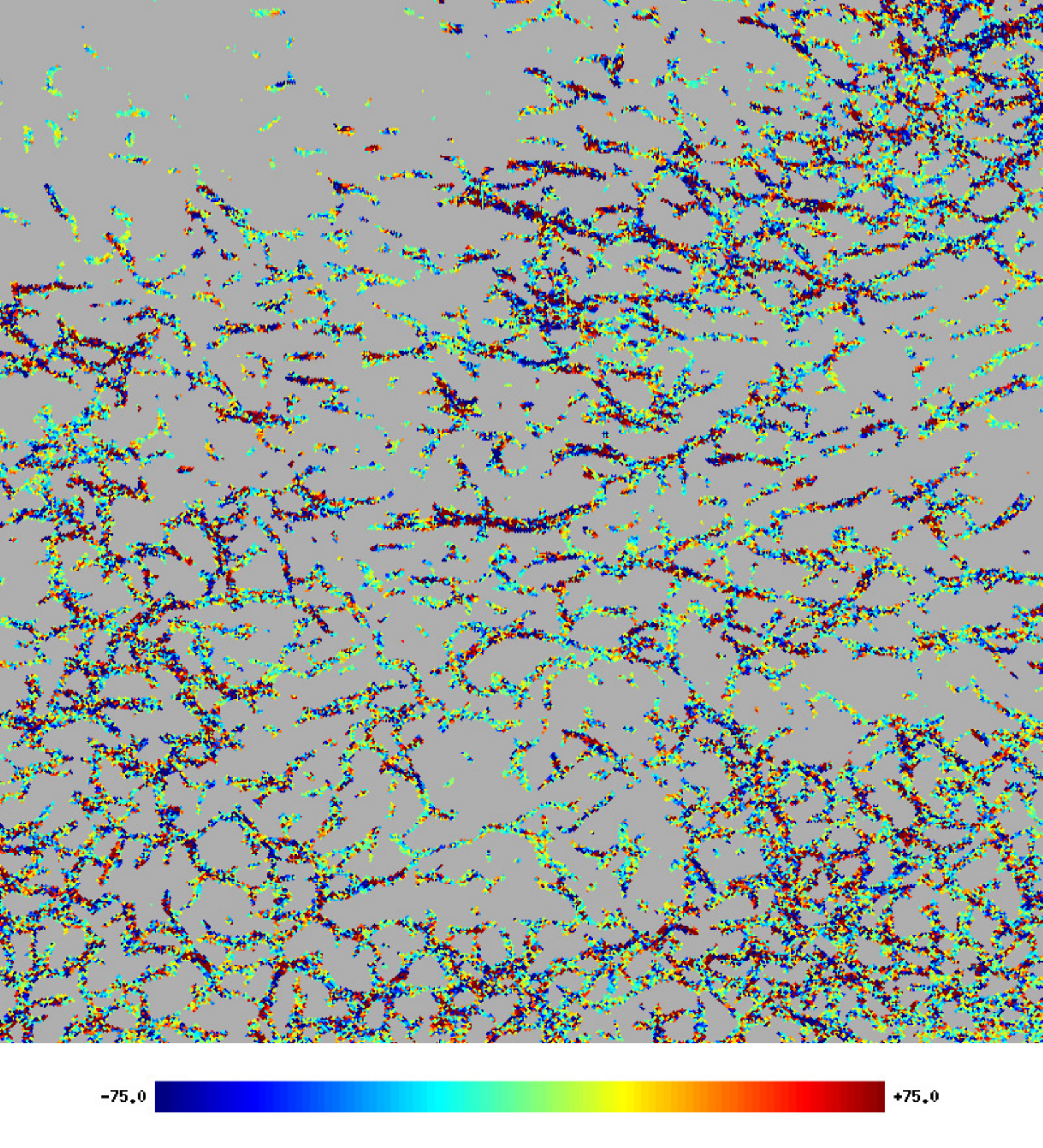}
   \caption{PPV$_\mathrm{fil}$ velocity gradient map according to 
     Eq. (\ref{eq:VGT_fil}) for comparison with
     Fig.   \ref{Fig_HI:diff}. 
   }
   \label{Fig_HI:diff_fil}
\end{figure}

\subsection{FIR filaments and the \hi\ haystack  }
\label{FIR_fil}

VCA is based on the analysis of velocity slices with variable thickness,
spanning   from thin to very broad velocity slices with
structures in column density, and claims to reconstruct the underlying
velocity and density characteristics. This approach fails, and as a
demonstration we want to compare a basic subset of Morse complexes with
very thick and thin velocity slices with FIR caustics. Figure
\ref{Fig_857_HI} first compares  FIR filaments with our derived
PPV$_\mathrm{fil}$ results and eigenvalues for \hi\ column densities
integrated over the range $ -50 < v_{\mathrm{LSR}} < 50 $
\kms. Instead, Fig.  \ref{Fig_HI} covers \hi\ eigenvalues
$\lambda_{-}$ for channel maps with a 1 \kms\ channel width at
velocities $ -4 < v_{\mathrm{LSR}} < 4 $ \kms. Fluctuations of the
eigenvalue strengths along the filaments are frequent.

\subsection{Filament tracing contra velocity slicing}
\label{tracing}

The FIR filaments in the left panel of Fig. \ref{Fig_857_HI} are
clearly well represented by a rather continuous PPV$_\mathrm{fil}$
distribution with a well-defined homogeneous velocity structure,
including some turbulent velocity fluctuations. This is the result from
\citetalias{Kalberla2021}, matching FIR and \hi\ structures with both
eigenvalues and orientation angles (from eigenvectors). The best fit
result was obtained by using local velocity slices with a thickness of 1
\kms.  Increasing the velocity slice thickness leads to a degradation of
the results.  Comparing FIR filaments (left in Fig. \ref{Fig_857_HI})
with filaments in column density eigenvalues (right), we find a very
complex distribution of filaments in column density. Only a small
fraction of the FIR structures can be traced in \hi\ column densities.

Matching FIR filaments with 1 \kms\ broad eigenvalue maps in
\hi\ visually is an even more complex task if PPV channel maps are
used. Trying to find individual FIR structures within these \hi\ maps in
Fig. \ref{Fig_HI} is like looking for needles in a haystack. The
\hi\ distribution is full of filaments. Only a small fraction of these
filaments are clearly linked to the FIR. Considering FIR filaments as flux
tubes, the filament curvature changes across the tube envelopes (see Sect. 5
of \citetalias{Kalberla2021}). Increased curvatures outside the filament
centers imply increased curling of magnetic field lines with decreasing
field strengths as predicted for the small-scale dynamo.

Comparing the individual maps in Fig. \ref{Fig_HI} with the FIR and
PPV$_\mathrm{fil}$ in Fig. \ref{Fig_857_HI} demonstrates impressively
the power of the Hessian operator as a search algorithm for structures
if both eigenvalues and eigenvectors (hence orientation angles) are
used to match the data. The implication is that \hi\ orientation angles are
strongly velocity dependent. Tracing FIR in $\theta$ recovers the
velocity structure of \hi\ caustics that remain otherwise hidden in
velocity slices (channel maps in PPV).

\subsection{Velocity gradients for caustics in PPV and PPV$_\mathrm{fil}$  }
\label{FIR_fil_grad}

Using caustics in PPV in place of PPV$_\mathrm{fil}$ databases has
significant consequences for the determination of velocity gradients. As
pointed out by \citet{Burkhart2021}, MHD theories predict that velocity
fluctuations trace magnetic field fluctuations around turbulent eddies. It
is expected that the amplitude of velocity gradients across these eddies
should increase with decreasing spatial scale. Thus, velocity gradients
are expected as a tracer of magnetic fields on small scales.

The velocity gradient technique (VGT) was first applied by
\citet{Gonzalez2017} to PPV data from an isothermal medium. Since
then, several publications used VGT with some modifications, we refer to
the detailed discussions in \cite{Hu2023}.  VGT needs to measure
gradients in PPV maps from velocity centroids or alternatively from
gradient angle maps by averaging over blocks with sizes around 100x100
pixels \citep{Lazarian2018}. This implies some smoothing. More importantly,  VGT does not use caustics to determine velocity gradients.

For the application of the Hessian operator only a 5x5 pixel kernel is
required. Caustics can be used then in a very simple way to derive
velocity gradients on small scales with an average filament width of
0.63 pc at a distance of 250 pc for HI4PI data. We define
\begin{equation}
  \frac{\partial \lambda_-(l,b,v_{\mathrm{LSR}})}{\partial v_{\mathrm{LSR}}} = \frac{\lambda_-(l,b,v_{\mathrm{LSR}} +
  \delta v_{\mathrm{LSR}} ) - \lambda_-(l,b,v_{\mathrm{LSR}} - 
  \delta v_{\mathrm{LSR}}) } {2~ \delta v_{\mathrm{LSR}}} 
        \label{eq:VGT}
,\end{equation}
and use the observational available channel width of $\delta
v_{\mathrm{LSR}} = 1$ \kms. Figure \ref{Fig_HI:diff} shows PPV gradient
maps for the same channels that are shown in
Fig. \ref{Fig_HI}. Comparing the two figures, it is obvious that velocity
gradients are small-scale structures, running strictly parallel to the
filament ridges.

Applying this gradient technique to the PPV$_\mathrm{fil}$ database
needs the modification
\begin{equation}
  \frac{\partial \lambda_-(l,b,v_{\mathrm{fil}})}{\partial v_{\mathrm{fil}}}
  = \frac{\lambda_-(l,b,v_{\mathrm{fil}} +
  \delta v_{\mathrm{LSR}} ) - \lambda_-(l,b,v_{\mathrm{fil}} - 
  \delta v_{\mathrm{LSR}}) } {2~ \delta v_{\mathrm{LSR}}} .
        \label{eq:VGT_fil}
\end{equation}
Figure \ref{Fig_HI:diff_fil} shows the result. The differences between 
Figs. \ref{Fig_HI:diff} and  \ref{Fig_HI:diff_fil} are striking, but are readily explained in the context of
caustics as descending manifolds. \hi\ filaments represent passes with a
narrow width in velocity space. PPV$_\mathrm{fil}$ filaments are located
on top of the passes. The Morse lemma tells us that for smooth functions
of variables with a nondegenerate critical point at the origin there
exits a local diffeomorphism that preserves the origin and approximates
the original function with a quadratic function
\citep[e.g.,][]{Castrigiano2004}. Thus, on top of a saddle it is flat, and
gradients are low. Figure \ref{Fig_HI:diff_fil} shows essentially only
noise, but no systematic gradients. As soon as we get to the wings of the
descending manifold, gradients may get steep. Backpackers, walking on
passes, know this rule by experience. Steep gradients with narrow
filamentary structures in Fig. \ref{Fig_HI:diff} come from the fact that
filaments in Fig. \ref{Fig_HI} are made of CNM, and hence they all have  
narrow line widths, typically $\Delta v_{\mathrm{LSR}} = 3$ \kms.

The filaments in Fig. \ref{Fig_HI} and the gradients in Fig. \ref{Fig_HI:diff}
are PPV structures, and hence they represent caustics that are affected by
turbulent velocity fluctuations. Orientation angles are known only for a
small fraction of these structures  associated with FIR filaments.
Only for this restricted subset of filamentary structures
(Fig. \ref{Fig_HI:diff_fil}) is it  possible to separate turbulent
contribution and average filament velocities as detailed in
Sect. \ref{Hessian}. These common FIR/\hi\ filaments represent prominent
ridges within a network of associated \hi\ filaments.

\section{Summary and conclusion }
\label{Summary}

\hi\ shows for large parts of the diffuse ISM filamentary structures and
it is well established that these are correlated with FIR emission and
aligned with the ambient magnetic field (e.g., \citealt{Clark2014},
\citealt{Clark2015} or \citealt{Kalberla2016}). These authors identified
the structures with actual \hi\ density filaments. The  interpretation of the filaments
as density structures is in conflict with \citet{Lazarian2000}. Their
VCA postulate is that turbulent velocity fluctuations along the line of
sight makes emitting elements at different distances from the observer
overlap in velocity space. This effect, also called velocity mapping,
causes velocity caustics that are said to be dealt with from a
mathematical perspective in \citet{Arnold1985}.\footnote[6]{We were
  unable to find a treatment of mapping effects, such as velocity
  crowding, in this book.}  Thus, multiple elements along the line of
sight at the same velocity cause enhanced emission in the observations
that does not exist in position space (\citealt{Lazarian2018} and Fig. 1
in \citealt{Hu2023}).

Following VCA, a number of publications corroborated the velocity caustics
model.  The discussion about density structures or velocity caustics
cumulated with the publications of \citet{Lazarian2018} and
\citet{Clark2019} in an open conflict. In turn, the existence of dense
structures and further the association between FIR structures and
\hi\ counterparts was proven (e.g., \citealt{Clark2019},
\citealt{Peek2019}, \citealt{Clark2019b}, \citealt{Murray2020},
\citealt{Kalberla2020}, \citetalias{Kalberla2021}, and
\citetalias{Kalberla2023}). Filaments were found to be associated with
particular cold CNM, an increased CNM fraction and an enhanced FIR
emissivity $I_{\mathrm{FIR}}/N_{\mathrm{HI}}$.   

In response, \citet{Yuen2021} developed a
numerical model  called VDA  to separate velocity and density
fluctuations in the VCA framework. This model was supposed to support
VCA, and this presentation has been recently reinforced by \citet{Hu2023}.

We applied the VDA model to HI4PI observations used by 
\citetalias{Kalberla2021} and \citetalias{Kalberla2023}. The main body
of our paper is about the performance of the VDA algorithm. 
Summarizing our results, we note first that structures that were called
velocity caustics in the literature are not caustics as defined by
\citet{Thom1975} and \citet{Arnold1985} or in the recent literature  by \citet{Sousbie2011a}, among others. VDA velocity and density data are modified
data products that, in a   way similar to that used for the original observations, may
contain caustics. These caustics need to be determined, following
the standard procedures explained   by \citet{Sousbie2011a},
\citet{Sousbie2011b}, \citet{Feldbrugge2018}, and 
\citet{Feldbrugge2019}, among others.

Since the year 2000 has become customary for  a group of authors to describe and
cite \hi\ structures as velocity caustics without giving any proof that
they are actually caustics. In the literature there exists a clear
  classification of critical points and procedures for deriving caustics,
  and one would expect   such definitions to be followed. Only a small
  fraction (typically 30\%) of the structures visible in VCA/VDA velocity
  caustic maps $p_v$ are related to generic caustics. These filamentary
  structures, classified as $A_3$ by \citet{Arnold1985}, must be worked
  out before they can be called caustics, otherwise the terminology is
  misleading. In other words, the caustics of the velocity caustics need to
  be determined. In the literature,  images of velocity caustics are published
  that do not at all show similarities to the examples of caustics from
the   observations given here. A few RHT images have been published  by
  \citet{Hu2023}, among others, but it is unclear how they are related to FIR
  structures. 

To determine caustics, we applied the same Hessian analysis as in
\citetalias{Kalberla2021} to the VDA data and found that caustics from
VDA velocity data are almost identical with caustics from the original
observations. The Pearson product-moment correlation coefficient is $
\ga 98\%$. Tracing {\it Plank} FIR caustics at 857 GHz, we recovered
97\% of the filaments from the original \hi\ database with the VDA
velocity data. In the mathematical framework of caustics in smooth
differentiable maps, caustics from VDA velocity data can be considered
as local diffeomorphisms to the caustics from direct
\hi\ observations. Thus, the VDA decomposition algorithm does not provide
any advantages. Comparing the alignment measures when fitting FIR and
\hi\ filament structures, the residual misalignment from fitting
uncertainties increases in case of VDA velocity caustics. We conclude
that \hi\ filament structures cannot be significantly affected by
turbulent velocity crowding, contrary to VCA/VDA predictions. The
density model based on original unmodified data performs better.
Filaments derived from VDA velocity caustics are correlated in the same
way  as caustics derived in
\citetalias{Kalberla2021}, with caustics from broadband FIR emission. The assertion by \citet{Hu2023} that the data
analysis in \citetalias{Kalberla2023} did not consider velocity caustics
is not justified (see \citealt{Kalberla2021b}, with a preliminary version
of Sect. \ref{VDA_caustics}).

In the same way as intensity fluctuations in the VDA framework can arise
from both density and velocity perturbations, VCA predicts that the
observed \hi\ power distribution can be decomposed in velocity and
density terms.  These can be characterized with narrow and broad
velocity channel data and, according to VCA, they should differ
significantly from each other. We calculated VDA velocity and density power spectra. The
power spectra are straight; there are no significant differences between
the VDA velocity power spectrum and the spectrum from the original
\hi\ data. All spectral indices agree within the errors. There is no
evidence for the VCA predicted change of the power spectrum with respect
to the used velocity width, defining velocity, or density effects in VDA.

The VDA predicts significant differences between data products from the original
\hi\ data and the decomposed velocity and density terms.  The result
that VDA velocity data and original \hi\ observations deliver nearly
identical caustics rises the question whether the VDA decomposition is
based on any physical background. One of the discussed topics is the
question whether filamentary \hi\ structures that originate from small-scale fluctuations are linked to the CNM with narrow line widths.

We considered the BIGHICAT compilation of absorption data that are
available from \citet{Naomi2023}. We found that  63 out of 66 observations without
detectable absorptions are located outside FIR/\hi filaments, and  174
positions from the absorption survey could be used to determine
absorption components within FIR/\hi filaments. The derived \hi\ spin
temperatures and also the corresponding volume densities are typical for
temperatures of the CNM. This is not a new result, but for the first time
it has been  shown that the CNM in the diffuse ISM is exclusively located in
filaments with FIR counterparts. We conclude that FIR/\hi\ filaments
trace the CNM in the same way as FIR filaments trace cold dust. These
caustics are density structures, consistent with previous
findings (e.g., \citealt{Clark2019}, \citealt{Peek2019},
\citealt{Clark2019b}, \citealt{Murray2020}, \citealt{Kalberla2020},
\citetalias{Kalberla2021}, and \citetalias{Kalberla2023}).

The VCA and descendants, as reviewed by \citet{Burkhart2021}, relies on a PPV
decomposition algorithm in the channel maps. We demonstrate in
Fig. \ref{Fig_HI} that individual channels in the PPV database are not
optimal for the detection and tracing of FIR filaments. In the  same way,
velocity gradients in PPV are hard to interpret
(Fig. \ref{Fig_HI:diff}). In \citetalias{Kalberla2021} it was shown that
the distribution of curvatures in filaments follows closely the expected
curvature distribution in the case of a turbulence-driven small-scale
dynamo. The filament curvature is related to the magnetic field strength
\citep{Schekochihin2004}. Sharply curved fields imply a high field
tension, and the field strength is reduced. For the small-scale
  dynamo the magnetic field amplification is exponentially fast and
  occurs due to stretching of the magnetic field lines by the random
  velocity shear associated with the turbulent eddies. According to
  \citet{Matthaeus2008}, the local directional alignment of the velocity
  and magnetic field fluctuations is expected to occur rapidly. However,
   these calculations missed an
  important term. The full expression is given by \citet{Soler2017},
  their MHD simulations, however, only cover    densities $ n \ga 3000 ~ {\rm
    cm^{-3}}$. Densities in our case are far below this value; we
  observe caustics and velocity gradients that are aligned with the
  magnetic field. In all cases \hi\ column densities are below $N_H =
  10^{21.7} \rm cm^{-3}$ and the magnetic field is expected to be
  aligned parallel to the filaments \citep{Hennebelle2019}. 

Observed local \hi\ orientation angles are strongly velocity
dependent. This implies that \hi\ filaments must have narrow velocity
dispersions. The turbulent velocity field is expected to introduce some
local filament bending. In projection, if this happens along the line of
side, the bending cannot be observed in position. However, a velocity
induced bending causes changes in orientation angle.  Filament tracing
is therefore a tracing in position on the plane of the sky and is analogous
in orientation angle for the velocity along the line of sight.
Instead of a PPV alignment we need to consider a
PPV$_\mathrm{fil}$ geometry, tracing the filament also in filament
velocity $v_\mathrm{fil}$. Such a tracing is provided in a natural way
if the Hessian analysis uses the full information, eigenvalues, and
eigenvectors (hence orientation angles) from FIR filaments. This
PPV$_\mathrm{fil}$ tracing has to be done for each individual filament
and $v_\mathrm{fil}$ includes position-dependent turbulent velocity
fluctuations. In the  case that filaments are positionally disjunct (at high
Galactic latitudes only a small amount of confusion is expected; see
\citetalias{Kalberla2023}) the $v_\mathrm{fil}$ data can be combined to
a large-scale velocity field for the filaments.

The significant difference between PPV$_\mathrm{fil}$ density structures
and PPV velocity crowding from the VCA theory is the distribution of
matter along the line of sight. In the first case fibers are bent as
density structures in PPV$_\mathrm{fil}$, in the second case the
distribution is fluffy along the line of sight because of different
separate components that cause \hi\ velocity crowding. Both approaches can only
be consistent if velocity crowding exists   on scales that are
compatible with the radial extensions of the FIR/\hi\ caustics. Such a
solution is however explicitly ruled out by \citet{Lazarian2018} for 
longwave-dominated power spectra with a steep spectral index ($\gamma
< -3$). A shortwave-dominated or shallow velocity spectrum with $\gamma
\ga -3 $ is not considered   to be physically motivated
\citep{Chepurnov2009}. 

Our results are in conflict with such theoretical expectations. We get
an average multiphase \hi\ spectral index of $\gamma = -2.85$. According
to \citet{Naomi2023}, this index is representative for the \hi\ at high
Galactic latitudes. For the CNM the power spectrum flattens to $\gamma =
-2.5$ \citep{Kalberla2019}, approaching $\gamma = -2.4$ as the spectral
index of the turbulent magnetic field \citep{Ghosh2017,Adak2020}.
Turbulence in FIR/\hi\ filaments is driven on small scales by a
fluctuation dynamo \citepalias{Kalberla2021}, in conflict with a longwave-dominated power law, as assumed by \citet{Lazarian2018}. Here we
argue based on observational results against theoretical
expectations.\footnote[7]{ \citet[][Sect. 7.3.6]{Hu2023} found that the
  decomposition of \hi\ observations by \citet{Kalberla2020b} result in
  steep CNM power spectra in the case of thin channels, suggesting that
  small-scale CNM structures are rare in thin channels compared to those
  in thick channels. Our   comment on this statement is the following:  Single
  channel CNM power spectra have been shown to be flat, $\gamma \ga
  -2.5$ (e.g., \citealt{Kalberla2019}, Figs. 1, 9, 13, and 23, and 
  \citealt{Kalberla2020b}, Figs. 3--5). The assertion that small-scale
  CNM structures are rarer in thin channels than in   thick
  channels is incompatible with \hi\ observations, also with our current
  results. Observational conflicts with \citet{Clark2019} that might
  support VCA as reported by \citet{Hu2023} in their Sect. 7.3.6 do not
  exist. }


\begin{acknowledgements}
  
We acknowledge the referee for careful reading and comments that
supported an improvement of the current manuscript.  We thank Naomi
McClure-Griffiths and Daniel Rybarczyk for help with the BIGHICAT
database.  We acknowledge the referee of a previous version of this
manuscript for constructive comments that helped to improve the quality
of this version as a separate paper. HI4PI is based on observations with
the 100-m telescope of the MPIfR (Max-Planck- Institut für
Radioastronomie) at Effelsberg and the Parkes Radio Telescope, which is
part of the Australia Telescope and is funded by the Commonwealth of
Australia for operation as a National Facility managed by CSIRO. This
research has made use of NASA's Astrophysics Data System.  Some of the
results in this paper have been derived using the HEALPix package.
   \end{acknowledgements}

\end{document}